\documentclass[useAMS,usenatbib]{mn2e}

\usepackage{graphicx}
\usepackage{amssymb}

\newcommand{\feh}{\mbox{[Fe/H]}}
\newcommand{\teff}{\mbox{$T_{\rm eff}$}}

\newcommand{\vsini}{\mbox{$v \sin I$}}

\newcommand{\kms}{\mbox{km\,s$^{-1}$}}
\newcommand{\ms}{\mbox{m\,s$^{-1}$}}

\newcommand{\mplanet}{\mbox{$M_{\rm pl}$}}
\newcommand{\rplanet}{\mbox{$R_{\rm pl}$}}
\newcommand{\densplanet}{\mbox{$\rho_{\rm pl}$}}
\newcommand{\mjup}{\mbox{$M_{\rm Jup}$}}
\newcommand{\rjup}{\mbox{$R_{\rm Jup}$}}
\newcommand{\densjup}{\mbox{$\rho_{\rm Jup}$}}
\newcommand{\mstar}{\mbox{$M_*$}}
\newcommand{\rstar}{\mbox{$R_*$}}
\newcommand{\densstar}{\mbox{$\rho_*$}}
\newcommand{\msol}{\mbox{$M_\odot$}}
\newcommand{\rsol}{\mbox{$R_\odot$}}
\newcommand{\denssol}{\mbox{$\rho_\odot$}}
\newcommand{\ecos}{\mbox{$e \cos \omega$}} 
\newcommand{\esin}{\mbox{$e \sin \omega$}} 
\def\secos{$\sqrt{e} \cos \omega$}
\def\sesin{$\sqrt{e} \sin \omega$}
\def\vsicos{$v \sin I \cos \lambda$}
\def\vsisin{$v \sin I \sin \lambda$}
\def\svsicos{$\sqrt{v \sin I} \cos \lambda$}
\def\svsisin{$\sqrt{v \sin I} \sin \lambda$}
\def\teql{$T_{\rm P,A=0}$}
\def\teqlf{$T_{\rm P,A=0,f}$}
\def\teqlfull{$T_{\rm P,A=0,f=1}$}
\def\teqlday{$T_{\rm P,A=0,f=2}$}
\def\ali{$A_{\rm Li}$}		

\def\fctwo{$\Delta F_{4.5\,\micron}$}
\def\fceight{$\Delta F_{8\,\micron}$}

\newcommand{\E}{\mathop{\bf E\/}}

\newcommand{\aap}{A\&A}
\newcommand{\apj}{ApJ}
\newcommand{\apjl}{ApJ}
\newcommand{\mnras}{MNRAS}
\newcommand{\apjs}{ApJS}
\newcommand{\nat}{Nature}
\newcommand{\pasp}{PASP}

\title[Thermal emission at 4.5 and 8 \micron\ of WASP-17b]
{Thermal emission at 4.5 and 8 \micron\ of WASP-17b, an extremely large planet 
in a slightly eccentric orbit}
\author[D.~R.~Anderson et al.]
{D.~R.~Anderson,$^{1}$\thanks{dra@astro.keele.ac.uk}
A.~M.~S.~Smith,$^{1}$
A.~A.~Lanotte,$^{2}$
T.~S.~Barman,$^{3}$
\newauthor
A.~Collier~Cameron,$^{4}$
C.~J.~Campo,$^{5}$
M.~Gillon,$^{2}$
J.~Harrington,$^{5}$
C.~Hellier,$^{1}$
\newauthor
P.~F.~L.~Maxted,$^{1}$
D.~Queloz,$^{6}$ 
A.~H.~M.~J.~Triaud,$^{6}$
P.~J.~Wheatley$^{7}$\\
$^{1}$Astrophysics Group, Keele University, Staffordshire ST5 5BG, UK\\
$^{2}$Institut d'Astrophysique et de G\'eophysique,  Universit\'e de Li\`ege,  All\'ee du 6 Ao\^ut, 17,  Bat.  B5C, Li\`ege 1, Belgium\\
$^{3}$Lowell Observatory, 1400 West Mars Hill Road, Flagstaff, AZ 86001, USA\\
$^{4}$SUPA, School of Physics and Astronomy, University of St. Andrews, North Haugh, Fife, KY16 9SS, UK\\
$^{5}$Planetary Sciences Group, Department of Physics, University of Central Florida, Orlando, FL 32816-2385, USA\\
$^{6}$Observatoire de Gen\`eve, Universit\'e de Gen\`eve, 51 Chemin des Maillettes, 1290 Sauverny, Switzerland\\
$^{7}$Department of Physics, University of Warwick, Coventry, CV4 7AL, UK}

\begin{document}

\date{Accepted 1988 December 15. Received 1988 December 14; in original form 1988 October 11}

\pagerange{\pageref{firstpage}--\pageref{lastpage}} \pubyear{2002}

\maketitle

\label{firstpage}

\begin{abstract}
We report the detection of thermal emission at 4.5 and 8 \micron\ from the 
planet WASP-17b. We used {\it Spitzer} to measure the system 
brightness at each wavelength during two occultations of the planet by its host 
star. 
By combining the resulting light curves with existing transit light curves and 
radial velocity measurements in a simultaneous analysis, we find   
the radius of WASP-17b to be 2.0 \rjup, which is 0.2 \rjup\ larger than any 
other known planet and 0.7 \rjup\ larger than predicted by the standard cooling 
theory of irradiated gas giant planets. 
We find the retrograde orbit of WASP-17b to be slightly eccentric, with  
$0.0012 < e < 0.070$ (3 $\sigma$). 
Such a low eccentricity suggests that, under current models, tidal heating alone 
could not have bloated the planet to its current size, so the radius of 
WASP-17b is currently unexplained. 
From the measured planet-star flux-density ratios we infer 4.5 and 8 \micron\ 
brightness temperatures of $1881 \pm 50$ K and $1580 \pm 150$ K, respectively, 
consistent with a low-albedo planet that efficiently redistributes heat from 
its day side to its night side. 
\end{abstract}


\section{Introduction}

WASP-17b is a transiting, 0.49-Jupiter-mass planet in a 3.74-day orbit around a 
metal-poor, 1.3-Solar-mass F6V star 
\citep[][hereafter A10]{2010ApJ...709..159A}. 
Data presented in A10, confirmed by \citet[][hereafter T10]{2010A&A...524A..25T} 
and \citet{2010ApJ...722L.224B}, indicated that WASP-17b is in a retrograde 
orbit around its host star, the first such orbit to be found. 

WASP-17b probably formed beyond the ice line in a near-circular, coplanar orbit, 
as predicted by the canonical model of star and planet formation. 
It may have subsequently acquired an eccentric, highly inclined orbit via 
planet-planet scattering \citep[e.g.][]{2008ApJ...686..621F}  
or the Kozai mechanism \citep[e.g.][]{2007ApJ...669.1298F, 2008ApJ...678..498N}. 
Tidal friction may then have shortened and circularised the long, eccentric 
orbit, with the energy being dissipated within the planet as heat 
\citep[e.g.][]{2008ApJ...681.1631J, 2009ApJ...700.1921I}. 

As the planet is low-mass and the star is hot (\teff\ = $6650 \pm 80 $K, T10) 
and fast-rotating (\vsini\ = 9.8 \kms, T10), the radial velocity (RV) 
measurements are relatively low signal-to-noise, and so the eccentricity of the 
planet's orbit was poorly constrained (e = 0--0.24) in the discovery paper 
(A10). 
This translated into uncertainties in the stellar radius (\rstar = 1.2--1.6 
\rsol) and thus in the planetary radius (\rplanet\ = 1.5--2.0 \rjup), 
meaning the planet is larger than predicted by the standard cooling theory of 
irradiated gas giant planets by 0.2--0.7 \rjup\ \citep{2007ApJ...659.1661F}.

Using a coupled radius-orbit evolutionary model, \citet{2009ApJ...700.1921I} 
demonstrated that planets can be inflated to radii of 2 \rjup\ and beyond during 
a transient phase of heating that accompanies the tidal circularisation of a 
short ($a \approx 0.1$ AU), highly eccentric ($e \approx 0.8$) orbit. 
Such a large radius persists for only a few tens of Myr, suggesting that we are 
unlikely to observe any one system during this brief stage. 
However, only 3--4 of the $\sim$100 known planets are extremely bloated and 
the transit technique does preferentially find large planets. 

\citet{2010A&A...516A..64L} argued that the tidal heating rate is underestimated 
for even moderately eccentric orbits in studies such as that of 
\citet{2009ApJ...700.1921I}. If true, then a large fraction of energy tidally 
dissipated within the planet would have been radiated away by the age typical 
of the most bloated planets (a few Gyr) and so could not have played a 
significant role in their observed bloating. \citet{2011ApJ...727...75I} admit 
that their equations are not applicable at large eccentricity, but counter that 
neither are those that \citet{2010A&A...516A..64L} use. They state that the 
current uncertainty in tidal theory means that no approach can be considered 
correct.

In A10, the derived radius of WASP-17b is largest (2.0 \rjup) when a circular 
orbit is imposed, smaller (1.7 \rjup) when eccentricity is a free parameter 
($e = 0.13$), and smaller again (1.5 \rjup) when a main-sequence prior is 
imposed on the star and eccentricity is let free ($e = 0.24$). 
\citet{2009ApJ...700.1921I} and \citet{2011ApJ...727...75I} each note that, 
compared to planets that did not undergo tidal heating, tidally inflated planets 
are still signicantly larger Gyr after their orbits have circularised and tidal 
heating has ceased. 
In each study though, the orbits are 
still significantly non-circular ($e \gtrsim 0.1$) when the planets are largest. 
Hence, if the orbit of WASP-17b were found to be near-circular (which would mean 
that \rplanet\ $\approx$ 2 \rjup) then it would seem unlikely that the planet 
could have been inflated by a single episode of tidal inflation.
Another possibility is an ongoing tidal heating scenario, as explored by 
\citet{2010ApJ...713..751I}, in which the orbit of WASP-17b would be kept 
non-circular by the continuing interaction with an as-yet undiscovered third 
body. 

In order to better constrain the stellar and planetary radii, the system age, 
and the potential transient and ongoing tidal heating rates, an improved 
determination of orbital eccentricity is required.
Further high-precision RV measurements were obtained and presented in T10.
These allowed eccentricity to be better constrained to 
$e = 0.066^{+0.030}_{-0.043}$. 
The best prospect of improving this situation further lay with the measurement 
of an occultation (as the planet passes behind the star), which would 
constrain \ecos, where $\omega$ is the argument of periastron. 

In addition, by observing occultations from  the ground 
\citep[e.g.][]{2010A&A...513L...3A} and from space 
\citep[e.g.][]{2010arXiv1004.0836W}, we are able to perform photometry and 
emission spectroscopy of exoplanets which are spatially unresolved from their 
host stars. 
This allows us to determine planet albedos and the rates at which energy is 
redistributed from the day side to the night side of the planet 
\citep[e.g.][]{2008ApJ...676L..61B}, and to infer the temperature 
structure \citep[e.g.][]{2008ApJ...673..526K} and chemical composition of 
planet atmospheres \citep[e.g.][]{2009ApJ...690L.114S}.

We present here observations of two occultations of WASP-17b, each of which was 
measured at both 4.5 and 8 \micron. We combine these 
new data with existing data in a simultaneous analysis to show that WASP-17b is 
the largest (\rplanet\ = 2.0 \rjup) and least-dense (\densplanet\ = 0.06 
\densjup) planet known, and is in a slightly eccentric orbit around a 2--3 
Gyr-old, F-type star. 
Exoplanet occultation photometry is at the limit of {\it Spitzer} systematics 
and reliable conclusions concerning atmospheres and orbits depend on a careful 
analysis of the data. We thus present a detailed description of our method. 

\section{New observations}

We observed two occultations of the planet WASP-17b by its $K_{s}=10.22$ host 
star with the {\it Spitzer Space Telescope} 
\citep{2004ApJS..154....1W} during 2009 April 24 and 2009 May 1.
On each date we measured the WASP-17 system with the Infrared Array Camera 
\citep[IRAC,][]{2004ApJS..154...10F} in full array mode 
($256\times256$\,pixels, 1.2~arcsec/pixel) simultaneously in channel 2 (4.5 
\micron) and channel 4 (8 \micron) for a duration of 8.4~hr.
We used an effective integration time of 10.4~s, resulting in 2319~images per 
channel per occultation. 
To reduce the time-dependent sensitivity of the 8~\micron\ channel 
\citep[e.g.][]{2008ApJ...673..526K}, we exposed the array to a bright, diffuse 
source (M42) for 214 frames immediately prior to each occultation observation.

We used the images calibrated by the standard {\it Spitzer} pipeline (version 
S18.7.0) and delivered to the community as Basic Calibrated Data (BCD). 
We added back to each image the estimated brightness of the zodiacal light in 
the sky dark \citep{2005PASP..117..978R} so photometric uncertainties and 
the optimal aperture radii are correctly determined. 
For each image, we converted flux from MJy~sr$^{-1}$ to electrons and then 
used {\sc iraf} to perform aperture photometry for WASP-17, using circular 
apertures with a range of radii (1--6~pixels). 
The apertures were centred by fitting a Gaussian profile on the target. 
The sky background was measured in an annulus extending from 8 to 12 pixels 
from the aperture centre, and was subtracted from the flux measured within the 
on-source apertures. 
We estimated the photometric uncertainty as the quadrature addition of the 
uncertainty in the sky background in the on-source aperture, 
the read-out noise, and the Poisson noise of the total background-subtracted 
counts within the on-source aperture.
We calculated the mid-exposure times in the HJD (UTC) time system from the 
MHJD\_OBS header values, which are the start times of the DCEs (Data Collection  
Events), by adding half the integration time (FRAMTIME) values.

\begin{figure*}
\begin{center}
$\begin{array}{ccc}
\includegraphics[scale=0.95]{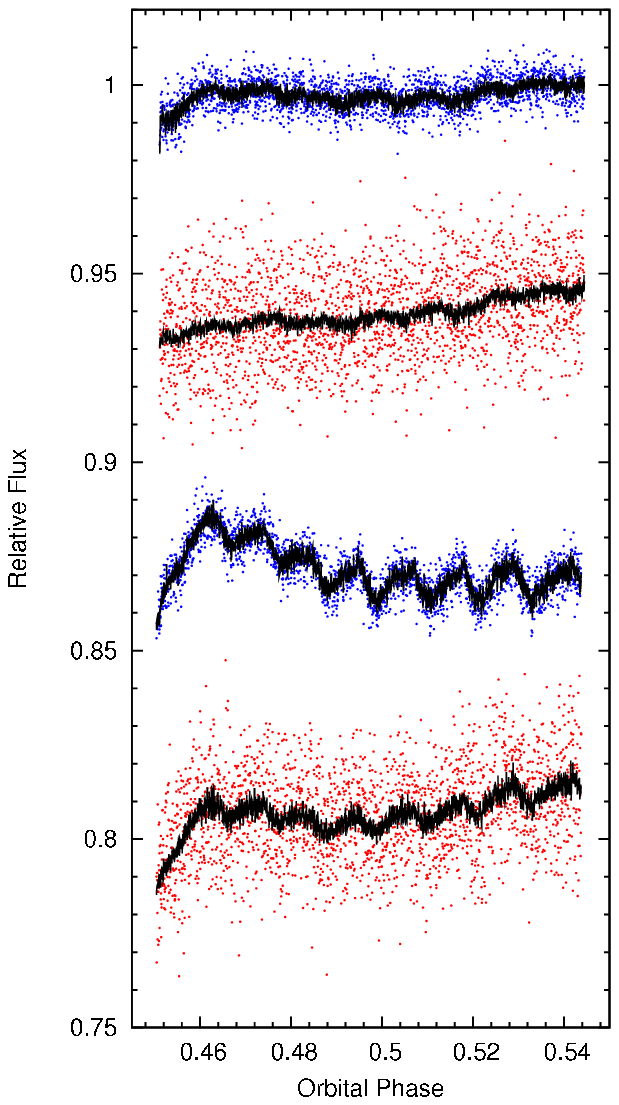} & 
\includegraphics[scale=0.95]{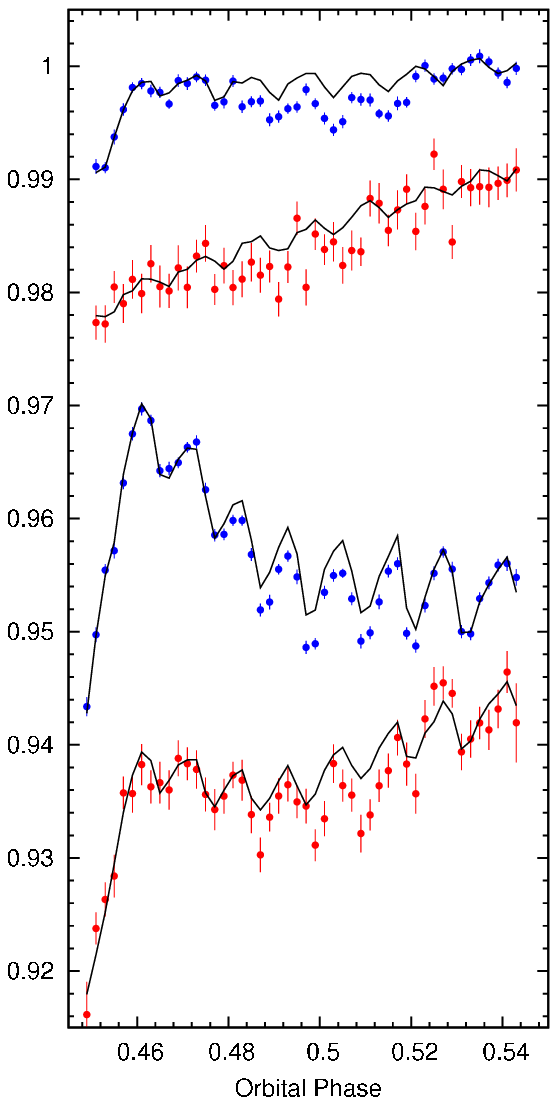} & 
\includegraphics[scale=0.95]{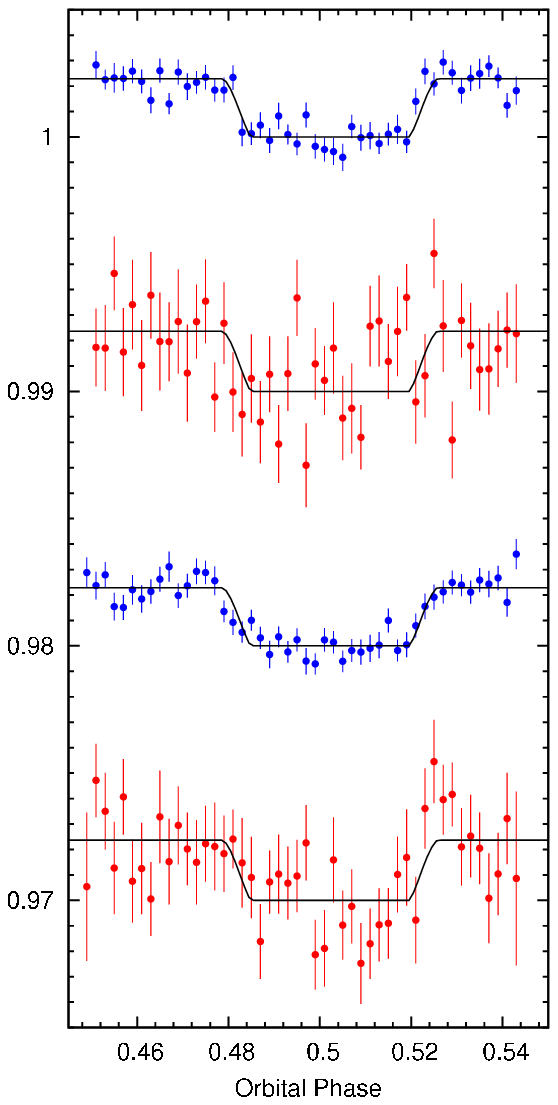}
\end{array}$
\caption{
In each of the above three plots, the upper two datasets were obtained at 4.5 
\micron\ (blue) and 8 \micron\ (red) on 2009 April 24 and the lower two 
datasets were taken at 4.5 \micron\ (blue) and 8 \micron\ (red) on 2009 May 
1. 
Relative flux offsets were applied to datasets for clarity. 
{\bf\em Left}: Raw {\it Spitzer} data with the best-fitting trend and 
occultation models superimposed.
{\bf\em Middle}: The same data binned in phase ($\Delta \phi=0.002, \sim$11 min) 
with the best-fitting trend models superimposed.
{\bf\em Right}: The binned data after dividing by the best-fitting trend models, 
and with the best-fitting occultation models superimposed.
We normalise the flux received from the star alone to unity, which is measured 
during occultation. 
\label{fig:spitz}}
\end{center}
\end{figure*}

We found (see Section 3) that for WASP-17 the highest signal-to-noise is 
obtained when using an aperture radius of 2.9 pixels for the 4.5 \micron\ data, 
and a radius of 1.6 pixels for the 8 \micron\ data. 
The data are displayed raw and binned in the first and second panels, 
respectively, of Figure~\ref{fig:spitz}.

We rejected any flux measurement that was discrepant with the median of its 20 
neighbours (a window width of 4.4 min) by more than four times its theoretical 
error bar.
We also performed a rejection on target position. For each image and for the 
{\it x} and {\it y} detector coordinates separately, we computed the difference 
between the fitted target 
position and the median of its 20 neighbours. For each dataset, we then 
calculated the standard deviation $\sigma$ of these {\it median differences} 
and rejected any points discrepant by more than 4 $\sigma$.
The numbers of points rejected on flux and target position for each dataset are 
given in Table~\ref{tab:rej}.

According to the IRAC handbook, each IRAC array receives approximately 1.5 
solar-proton and cosmic-ray hits per second, with $\sim$2 pixels affected in 
channel 2, and $\sim$6 pixels per hit affected in channel 4, while the cosmic 
ray flux varies randomly by up to a factor of a few over time-scales of minutes. 
Thus the probability per exposure that pixels within the stellar aperture will 
be affected by a cosmic ray hit is 1.5 per cent for channel 2 and 1.3 per cent 
for channel 4, which is in good agreement with the small portion of frames that 
we rejected. 

\begin{table}
\begin{center}
\caption{Number of points rejected per dataset per criterion.\label{tab:rej}}
\begin{tabular}{lrrrr}
\hline
\hline
Dataset			& Flux	& {\it x}-pos	& {\it y}-pos	& Total (per cent)\\
\hline
2009 Apr 24 / 4.5 \micron	& 34	&  4	& 1		& 35 (1.5)\\
2009 Apr 24 / 8 \micron		& 29	& 31	& 30		& 45 (1.9)\\
2009 May ~1 / 4.5 \micron	& 35	&  6	&  6		& 40 (1.7)\\
2009 May ~1 / 8 \micron		& 17	& 25	& 22		& 37 (1.6)\\
\hline
\end{tabular}
\end{center}
\end{table}

\section{Data analysis}

\subsection{Data and model}
We determined the system parameters from a simultaneous analysis incorporating: 
our new {\it Spitzer} occultation photometry; the WASP discovery photometry 
covering  the full orbit for the three seasons (March to August) of 2006--2008 
and presented in A10; 
a high-precision, {\it I$_{c}$}-band transit light curve taken with the 
1.2-m Euler-Swiss telescope on 2008 May 6 and presented in A10; 
and 124 RV measurements, including 34 taken during transit, 
made with the CORALIE and HARPS spectrographs and presented in A10 and T10.

These data were input into an adaptive Markov-chain Monte Carlo (MCMC) 
algorithm \citep{2007MNRAS.380.1230C,2008MNRAS.385.1576P, 2010A&A...516A..33E}. 
Such a simultaneous analysis is necessary to take account of the 
cross-dependency of system parameters and to make an honest assessment of their 
uncertainties. 
We used the following as MCMC proposal parameters: $T_{\rm c}$, $P$, $\Delta F$, 
$T_{14}$, $b$, $K_{\rm 1}$, \teff, \feh, \secos, \sesin, \svsicos, \svsisin, 
\fctwo\ and \fceight\ (see Table~\ref{tab:sys-params} for definitions). 

At each step in the MCMC procedure, each proposal parameter is perturbed from
its previous value by a small, random amount. 
Stellar density, which is constrained by the shape of the transit light curve 
\citep{2003ApJ...585.1038S} and the eccentricity of the orbit, is calculated 
from the proposal parameter values. This is input, together with the latest 
values of \teff\ and \feh\ (which are controlled by Gaussian priors) into the 
empirical mass calibration of \citet{2010A&A...516A..33E} to obtain an estimate 
of the stellar mass \mstar. 
From the proposal parameters, model light and RV curves are generated and 
$\chi^{2}$ is calculated from their comparison with the data. 
A step is accepted if $\chi^{2}$ (our merit function) is lower than for the 
previous step, and a step with higher $\chi^{2}$ is accepted with probability 
$\exp(-\Delta \chi^{2}/2)$. 
In this way, the parameter space around the optimum solution is thoroughly 
explored. 
The value and uncertainty of each parameter are respectively taken as the 
median and central 68.3 per cent confidence interval of the parameter's 
marginalised posterior probability distribution.

As \citet{2006ApJ...642..505F} notes, it is convenient to use 
\ecos\ and \esin\ as MCMC proposal parameters, because these two 
quantities are nearly orthogonal and their joint probability density function is 
well-behaved when the eccentricity is small and $\omega$ is highly uncertain. 
Ford cautions, however, that the use of \ecos\ and \esin\ as proposal parameters
implicitly imposes a prior on the eccentricity that increases linearly with $e$.
As such, we instead use \secos\ and \sesin\ as proposal parameters, which
restores a uniform prior on $e$. For similar reasons, we use \svsicos\ and 
\svsisin\ rather than \vsicos\ and \vsisin\ to parameterise the 
Rossiter-McLaughlin effect \citep[e.g.][]{2007ApJ...655..550G}. 

\subsection{Spitzer data}

\subsubsection{Deciding between models and datasets}
Systematics are present in IRAC photometry at a level similar to 
the predicted planetary occultation signal. Therefore, it is necessary to 
carefully detrend the photometry so as to obtain accurate occultation depths and 
timings.
To discriminate between various detrending models we used the Bayesian 
Information Criterion \citep[BIC,][]{schwarz197801}:

\begin{equation}
\label{eqn:bic}
{\rm BIC} = \chi^2 + k \ln N
\end{equation} 

\noindent where $k$ is the number of free model parameters and $N$ is the number 
of data-points. The BIC prefers simpler models unless the addition of extra 
terms significantly improves the fit. As such, it is a useful tool for selecting 
between models with different numbers of free parameters.

In addition, we used the root mean square (RMS) of the residuals about the 
best-fitting trend and occultation models to discriminate between different 
light curves obtained from different reductions of the same sets of images.

\subsubsection{Aperture radii}

\begin{figure}
\includegraphics{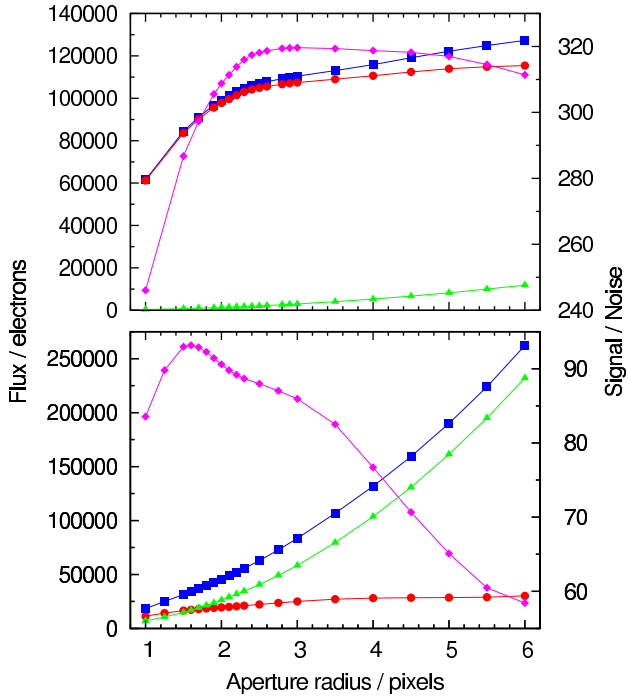}
\caption{
The flux due to WASP-17 (red circles), the sky and instrumental background 
(green triangles), and both WASP-17 and the background combined (blue squares), 
as well as the signal-to-noise (magenta diamonds) as a function of aperture 
radius. 
The upper and lower panels show the 4.5 \micron\ and 8 \micron\ data 
respectively. We show the data from 2009 Apr 24, but the data from 
2009 May 1 produce near-identical plots. 
\label{fig:star-sky}}
\end{figure}

We determined the optimal aperture radii to use for the 4.5 and 8~\micron\ data 
by performing aperture photometry with a range of aperture radii (1--6 pixels), 
and choosing the radii that produced the maximal signal-to-noise  
\citep[Figure~\ref{fig:star-sky}; e.g.][]{1989PASP..101..616H}. 
 
For the 4.5~\micron\ data, the radius that results in the highest 
signal-to-noise is 2.9 pixels and it is this radius that we adopt 
(Figure \ref{fig:star-sky}, upper panel). This radius incorporates the majority 
($\sim$92.5 per cent) of the target flux and little background flux ($\sim$2.6 
per cent of that of the target). 

At 8~\micron, as compared to 4.5~\micron, the background is brighter by a 
factor $\sim$20 and the source is fainter by a factor $\sim$4. 
Thus, with increasing aperture radius, the background flux 
quickly dominates the target flux (Figure~\ref{fig:star-sky}, lower panel). 
Indeed, the background flux equals the target flux when using our adopted, 
optimal (highest signal-to-noise) radius of only 1.6 pixels. 
This radius incorporates $\sim$60 per cent of the target flux and a similar 
amount (97 per cent of the target) of background flux.
The background flux is greater than the target flux by factors of approximately 
1.5, 2, 3 and 7.5 within apertures with radii of 2.2, 2.75, 3.5 and 6 pixels 
respectively. 

\subsubsection{Systematics}
IRAC uses an InSb detector to detect light around 4.5 \micron, and the 
measured flux exhibits a strong correlation with the position of the target 
star on the array. 
This effect is due to the inhomogeneous intra-pixel sensitivity of the detector 
and is well-documented 
\citep[e.g.][and references therein]{2008ApJ...673..526K}.
Following \citet{2008ApJ...686.1341C} we modelled this effect as a 
quadratic function of the sub-pixel position of the PSF centre, 
but with the addition of a linear term in time:

\begin{equation}
\label{eqn:ch2}
df = a_0 + a_xdx + a_ydy + a_{xx}dx^2 + a_{yy}dy^2 + a_tdt
\end{equation}

\noindent where $df = f - \hat{f}$ is the stellar flux relative to its weighted 
mean, $dx = x - \hat{x}$ and $dy = y - \hat{y}$ are the coordinates of the PSF 
centre relative to their weighted means, $dt$ is the time since the 
beginning of the observation, and $a_0$, $a_x$, $a_y$, $a_{xx}$, $a_{yy}$ and 
$a_t$ are coefficients.
We determined the trend model coefficients by linear least-squares minimization 
at each MCMC step, subsequent to division of the data by the eclipse model. 
We used singular value decomposition 
\citep{press_numerical_1992} for this purpose.
Though a common eclipse model was fitted to occultation data from the same 
channel, trend models were fitted separately to each dataset.

The best-fitting trend models are superimposed on the binned photometry in the 
middle panel (first and third curves from the top) of Figure~\ref{fig:spitz}. 
Table~\ref{tab:coeffs} gives the best-fitting values for the trend model 
parameters (Equation~\ref{eqn:ch2}), together with their 1-$\sigma$ 
uncertainties.  

We found consistent 4.5 \micron\ eclipse depths when incorporating one of the 
two datasets or both of them in our analysis: 
$\Delta F_{{4.5\,\micron}_1} = 0.00225 \pm 0.00015$, 
$\Delta F_{{4.5\,\micron}_2} = 0.00244 \pm 0.00020$, 
and $\Delta F_{4.5\,\micron_{1+2}}   = 0.00230 \pm 0.00012$.

\begin{table}
\caption{Trend model parameters and coefficients} 
\label{tab:coeffs} 
\begin{tabular*}{0.5\textwidth}{@{\extracolsep{\fill}}lcccc} 
\hline 
 & \multicolumn{2}{c}{4.5 \micron} & \multicolumn{2}{c}{8 \micron} \\ 
 & 2009 Apr 24 & 2009 May 1 & 2009 Apr 24 & 2009 May 1 \\
\hline
\\
$\hat{f}$	& 106955.09		& 107395.47			& 17289.49			& 17866.66			\\
$\hat{x}$	& 24.28			& 24.85				& 25.23				& 25.64				\\
$\hat{y}$	& 25.28			& 25.44				& 23.02				& 23.18				\\
$a_0$		& $-26.0^{+3.6}_{-3.8}$		& $-10.8^{+10.7}_{-10.4}$	& $-108.42^{+0.67}_{-0.77}$	& $-94.2^{+3.1}_{-3.1}$		\\
$a_x$		& $-2606.7^{+13.1}_{-13.3}$	& 4927.2$^{+61.1}_{-60.5}$	& 720.2$^{+6.5}_{-6.2}$		& $-142.0^{+19.1}_{-20.1}$	\\
$a_y$		& $-5049.5^{+20.2}_{-20.3}$	& $-6866.5^{+11.9}_{-11.6}$	& 42.3$^{+5.2}_{-4.9}$		& $-1301.6^{+7.7}_{-7.2}$	\\
$a_{xx}$	& 8345.6$^{+479.7}_{-487.6}$	& $-105.0^{+636.2}_{-661.9}$	& $-1368.7^{+114.6}_{-110.7}$	& $-1748.4^{+401.7}_{-398.4}$	\\
$a_{yy}$	& $-8340.8^{+385.5}_{-416.0}$	& 3663.4$^{+155.7}_{-165.5}$	& $-1246.8^{+136.0}_{-134.3}$	& 905.4$^{+122.2}_{-121.2}$	\\
$a_t$		& 207.7$^{+20.7}_{-19.2}$	& $-14.5^{+35.8}_{-37.0}$	& 648.9$^{+5.2}_{-4.9}$		& 557.7$^{+9.0}_{-9.1}$		\\
\\
\hline
\end{tabular*}
\end{table}

\begin{figure*}
\begin{center}
$\begin{array}{cc}
\includegraphics[scale=1.2]{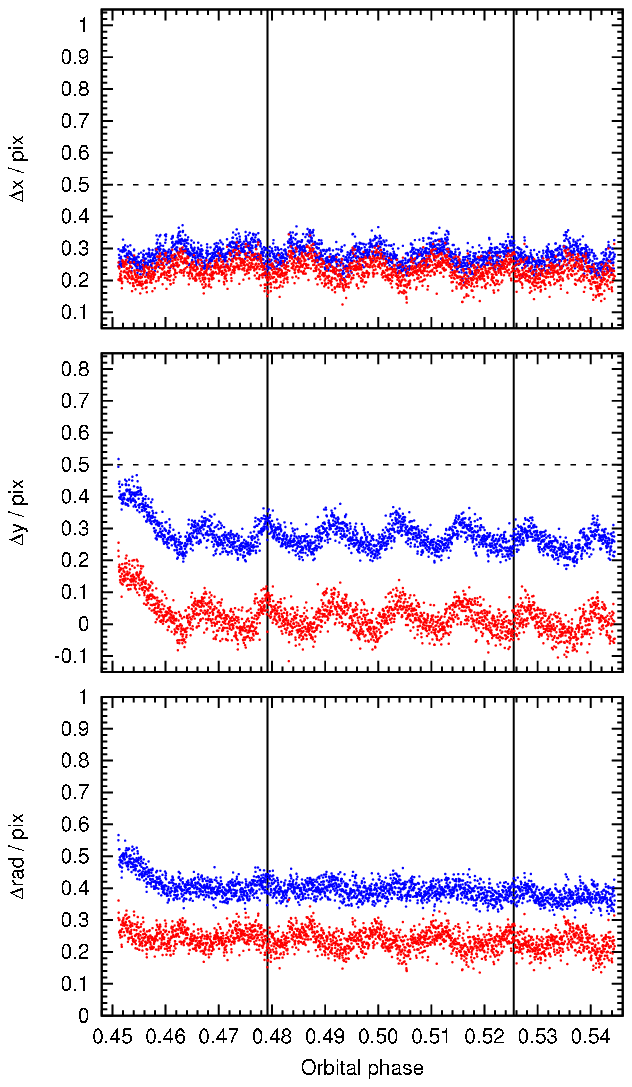} &
\includegraphics[scale=1.2]{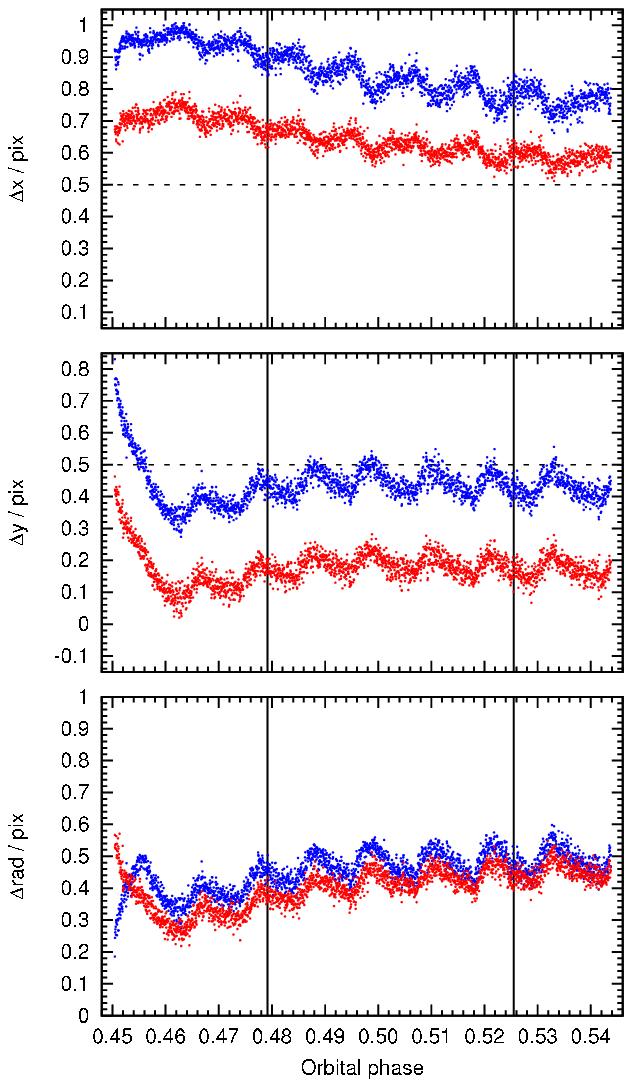}
\end{array}$
\caption{The detector positions of WASP-17's PSF centre for the first 
occultation (left-hand plots) and the second occultation (right-hand plots). 
The PSF centre positions on the 4.5 and 8 \micron\ detectors are 
respectively depicted by blue and red dots.
For each occultation we show the distance of the PSF centre from the nearest 
pixel centre in the $x$ and $y$ directions (top and middle panels 
respectively) and in the radial direction (bottom panels). 
Pixel centres are located at ($x$, $y$) = (0, 0) while pixel edges are located 
at (0.5, 0.5) and are demarcated by dashed lines.
\label{fig:eclpos}}
\end{center}
\end{figure*}

The systematics in the data from 2009 April 24 are of much smaller amplitude 
than those in the data from 2009 May 1. This is due to a chance placement of 
the target star on the detector. 
The detector positions of WASP-17's PSF centre are shown in 
Figure~\ref{fig:eclpos}. 
During each occultation and in each channel the motion due to the nodding of 
the spacecraft is evident in the $x$ and $y$ positions of the PSF centres. 
However, contrasting the two occultations, there is a marked difference in the 
radial distance from the nearest pixel centre over the course of the 
observations. 
In the data from 2009 Apr 24 the placement of the target on the 
detectors is such that the motion of the spacecraft in the $x$-direction 
largely compensates for the motion in the $y$-direction, resulting in a near 
constant radial distance from the nearest pixel centre. 
The opposite is the case in the 2009 May 1 dataset, where the motion in the 
$x$ and $y$ directions combines to produce large-amplitude oscillations in the 
distance from the nearest pixel centre. 
This results in the large saw-tooth systematics seen in the light curve (second 
panel of Figure~\ref{fig:spitz}).

Near the beginning of the observations on 2009 May 1, the target crossed a pixel 
boundary on the 4.5 \micron\ detector (Figure~\ref{fig:eclpos}, middle-right 
plot). 
This resulted in a point of inflection in the distance of the target from the 
nearest pixel centre (Figure~\ref{fig:eclpos}, bottom-right plot). 
As the sensitivity is higher toward the pixel centre and lower near the edges, 
it is therefore curious that no corresponding inflection point is seen in the 
light curve (Figure~\ref{fig:spitz}, middle panel).

IRAC uses a SiAs detector to observe at 8 \micron, and its response is 
usually thought to be homogeneous, though another systematic affects the 
photometry. This effect is known as the `ramp' because it causes the gain to 
increase asymptotically over time for every pixel, with an amplitude depending 
on a pixel's illumination history 
\citep[e.g.][and references therein]{2008ApJ...673..526K}. 
Again following \citet{2008ApJ...686.1341C}, we modelled this ramp as a 
quadratic function of $\ln(dt)$:

\begin{equation}
\label{eqn:ch4}
df = a_0 + a_1\ln(dt+t_{\rm off}) + a_2(\ln(dt+t_{\rm off}))^2
\end{equation}

\noindent where $t_{\rm off}$ is a proposal parameter. To prevent $t_{\rm off}$ 
from drifting more than an hour or so prior to the first observation, we place 
on it a Gaussian prior by adding a Bayesian penalty to our merit function 
($\chi^2$):

\begin{equation}
BP_{t_{\rm off}} = t_{\rm off}^2 / \sigma_{t_{\rm off}}^2
\end{equation}

\noindent where $\sigma_{t_{\rm off}}$ = 15 min.

\begin{figure*}
\begin{center}
$\begin{array}{ccc}
\includegraphics[scale=0.95]{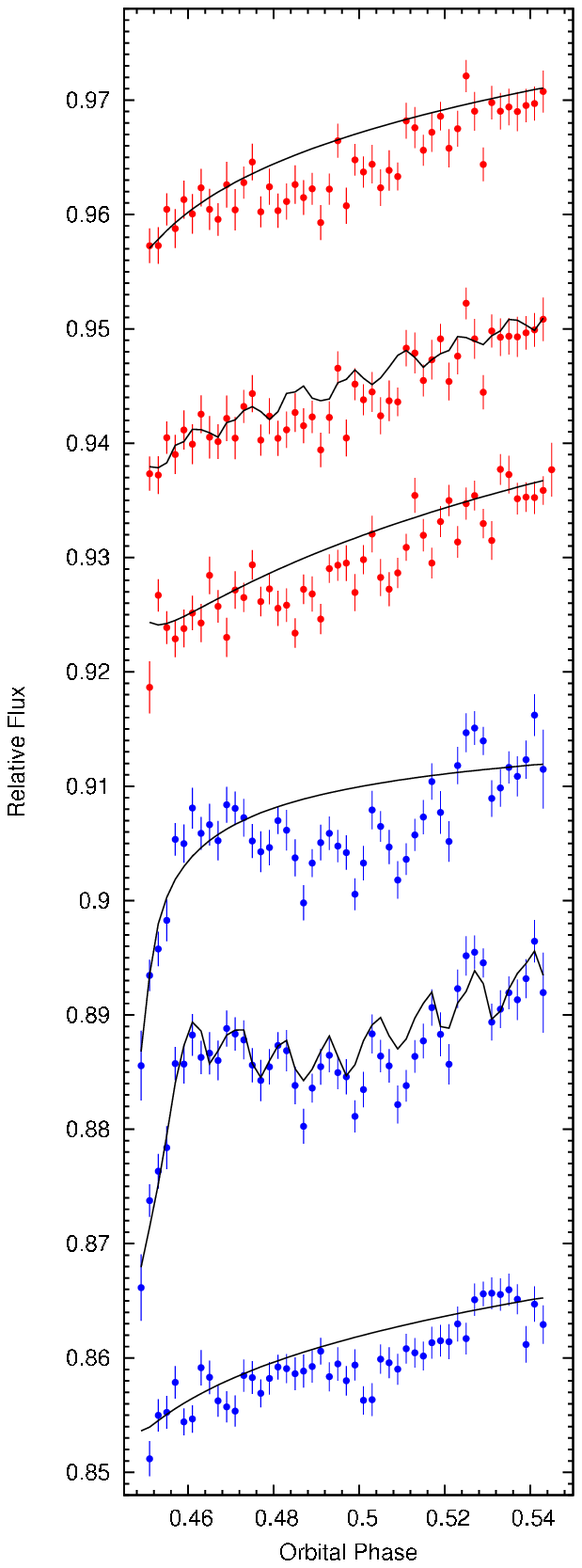} & 
\includegraphics[scale=0.95]{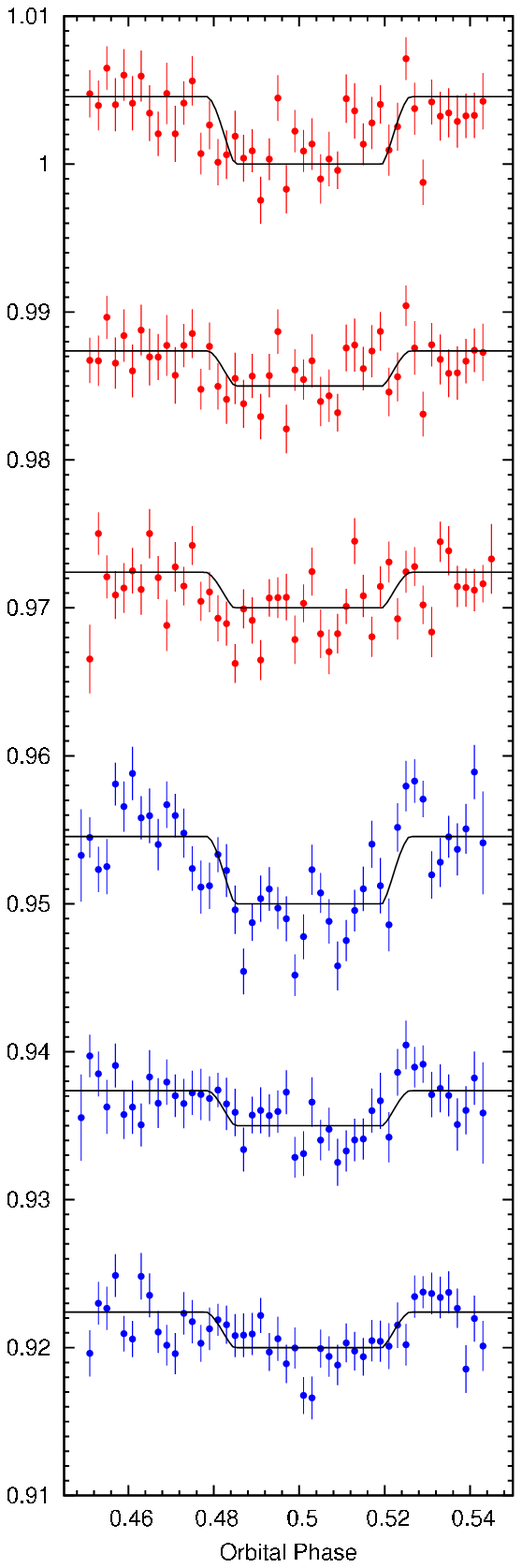} &
\includegraphics[scale=0.95]{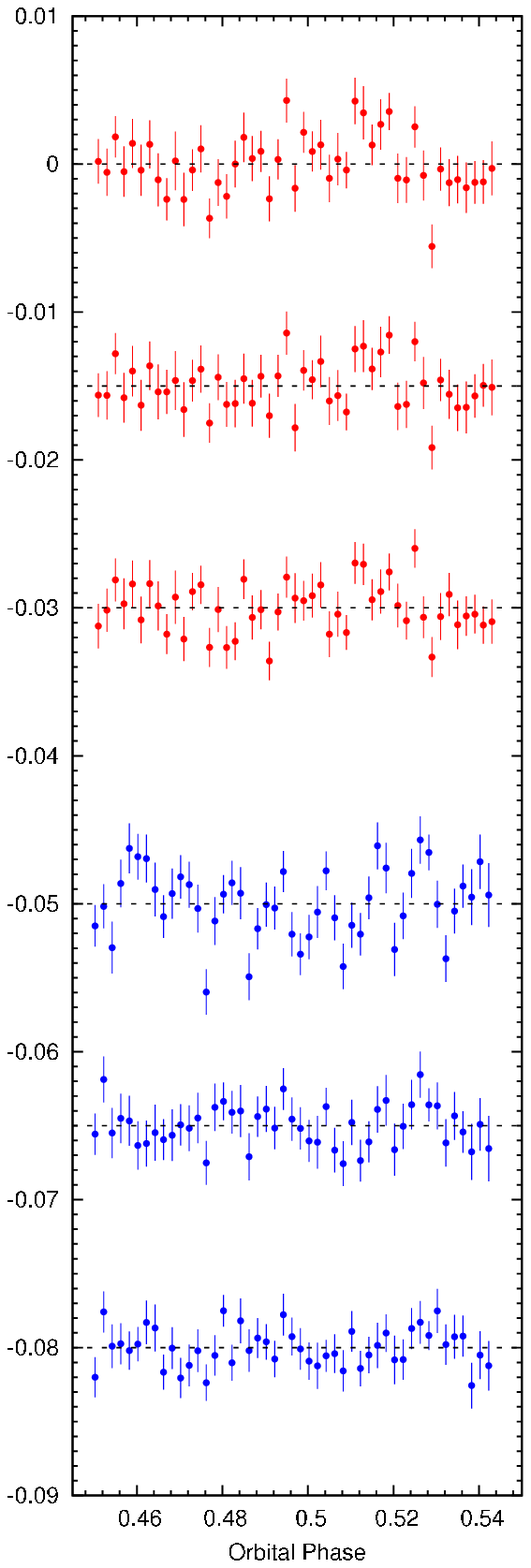}
\end{array}$
\caption{
A comparison of the methods for reducing and detrending the 8 \micron\ data. 
In each of the above three panels, the upper three curves (red dots) are the 
data from 2009 Apr 24 and the lower three curves (blue dots) are the data from 
2009 May 1. 
In each triplet of curves, the top curve is the light curve obtained by 
aperture photometry and detrended with Equation \ref{eqn:ch4}. The middle curve 
is the light curve obtained by aperture photometry and detrended with 
Equation \ref{eqn:ch2}. The bottom curve is the light curve obtained by 
deconvolution photometry and detrended with Equation \ref{eqn:ch4}.
{\bf\em Left panel}: Binned raw data, with the best-fitting trend models 
superimposed. 
{\bf\em Middle panel}: Binned detrended data, with the best-fitting occultation 
models superimposed. 
{\bf\em Right panel}: Binned residuals about the best-fitting trend and 
occultation models. 
\label{fig:spitz-ch4}}
\end{center}
\end{figure*}

From an initial MCMC run, we observed systematics in the residuals of the second 
8 \micron\ dataset, and so investigated decorrelating the 8 \micron\ data with 
detector position. 
A significantly lower occultation BIC ($\Delta$BIC = $-141$) resulted when 
also detrending for detector position, i.e. detrending with 
Equation~\ref{eqn:ch2} rather than with Equation~\ref{eqn:ch4}.
In addition, there was less scatter in the 8 \micron\ data when decorrelating 
with detector position (i.e. when detrending with Equation~\ref{eqn:ch2} rather 
than with Equation~\ref{eqn:ch4}; Figure~\ref{fig:spitz-ch4} and 
Table~\ref{tab:decphot-depths}). 
When not decorrelating with detector position, significantly different 
best-fitting 8 \micron\ occultation depths were obtained for the two individual 
datasets (Table~\ref{tab:decphot-depths}) and the depth obtained from the 
combined datasets was much deeper than otherwise. 
For these reasons and for reasons that will be presented in the remainder 
of this section, we opted to decorrelate the 8 \micron\ data with detector 
position. 

\begin{table*}
\caption{A comparison of the 8 \micron\ occultation depths and residuals 
from deconvolution photometry and aperture photometry.
\label{tab:decphot-depths}}
\begin{tabular}{llllllll}
\hline
Method		& Trend eq.	&$\Delta F_{{8\,\micron}_{1+2}}$& $\Delta F_{{8\,\micron}_1}$	& $\Delta F_{8\,\micron_2}$	&RMS$_{8 \micron_{1+2}}$& RMS$_{8 \micron_1}$	& RMS$_{8 \micron_2}$	\\
\hline
aper. phot.	& \ref{eqn:ch2} & $0.00238 \pm 0.00036$		& $0.00187 \pm 0.00050$		& $0.00303 \pm 0.00059$		& 0.01103		& 0.01142		& 0.01010		\\
aper. phot.	& \ref{eqn:ch4}	& $0.00455 \pm 0.00036$		& $0.00246 \pm 0.00058$		& $0.00643 \pm 0.00055$		& 0.01112		& 0.01127		& 0.01119		\\
decon. phot.	& \ref{eqn:ch4}	& $0.00240 \pm 0.00036$		& $0.00202 \pm 0.00052$		& $0.00276 \pm 0.00047$		& 0.00994		& 0.01019		& 0.00967		\\
\hline
\end{tabular}
\end{table*}

In Section~\ref{sec:decon} we use deconvolution photometry to show that the 
observed dependence on detector position is likely to have been introduced 
during aperture photometry. Therefore there is no evidence of an inhomogeneous 
intrapixel response of the SiAs 8 \micron\ detector, contrary to the case with 
the InSb 4.5 \micron\ detector.

The second and fourth curves from the top in the middle panel of 
Figure~\ref{fig:spitz} are the best-fitting trend models when detrending the 
two 8 \micron\ datasets with Equation~\ref{eqn:ch2}.
Note that the saw-tooth patterns of the 8 \micron\ trend models are in phase 
with those of the 4.5 \micron\ trend models, though each dataset was fit 
separately with its own trend model. 

In addition to Equation~\ref{eqn:ch2}, which we will call {\sc linear time}, 
we tried trend functions with a variety of time-dependency. 
These were {\sc no time}:

\begin{equation}
\label{eqn:not}
df = spatial
\end{equation}

\noindent where $spatial = a_0 + a_xdx + a_ydy + a_{xx}dx^2 + a_{yy}dy^2$ 
represents the detector position terms and an offset; 
{\sc quad time}:

\begin{equation}
\label{eqn:quadt}
df = spatial + a_tdt + a_{tt}dt^2;
\end{equation}

\noindent{\sc linear ln time}:

\begin{equation}
\label{eqn:linlnt}
df = spatial + a_1\ln(dt+t_{\rm off});
\end{equation}

\noindent{\sc quad ln time}:

\begin{equation}
\label{eqn:quadlnt}
df = spatial + a_1\ln(dt+t_{\rm off}) + a_2(\ln(dt+t_{\rm off}))^2;
\end{equation}

\noindent and {\sc rising exp} \citep{2007Natur.447..691H}:

\begin{equation}
\label{eqn:risexp}
df = spatial + a_3\exp(a_4dt).
\end{equation}

\noindent where $a_4$ is a proposal parameter. 

In Table~\ref{tab:ch4-depths-bics} we present the occultation depths and BIC 
values resulting from detrending the 8 \micron\ data with the various models. 
The 4.5 and 8 \micron\ data were fitted simultaneously, so an improved 
fit to the 8 \micron\ data would not be preferred if the fit to the 4.5 
\micron\ data were considerably worse. 
In Table~\ref{tab:ch4-depths-bics} the models are presented in descending order 
by how well they fit the combined 8 \micron\ datasets, and the BIC values are 
given relative to the best-fitting model ({\sc linear time}). 
The {\sc linear time} model is strongly favoured and we thus adopt this as our 
trend model for the 8 \micron\ data. 
The BIC values resulting from MCMC runs incorporating the two 8 \micron\ 
datasets individually ($8 \micron_1$ and $8 \micron_2$) are also given, with a 
similar order of preference to that of the combined datasets. Again, the 
{\sc linear time} model is clearly favoured over the others, supporting our 
decision to use the same model for the two datasets.
The occultation depths are consistent between the four most preferred models, 
but are not so for the two less preferred models. 
The depths found from the first dataset are shallower than those found from the 
second dataset, with a difference between the two of $1.7 \sigma$ in the case 
of the {\sc linear time} trend model. 

\begin{table*}
\caption{The 8-\micron\ occultation depths and the combined (4.5 and 8 \micron) 
relative occultation BIC values, when detrending the 8 \micron\ data with the 
various models 
\label{tab:ch4-depths-bics}}
\begin{tabular}{lllllrrr}
\hline
Model			& Eq.   & $\Delta F_{8\,\micron_{1+2}}$   & $\Delta F_{8\,\micron_{1}}$ & $\Delta F_{8\,\micron_{2}}$ & $\Delta$BIC$_{8 \micron_{1+2}}$ & $\Delta$BIC$_{8 \micron_1}$ & $\Delta$BIC$_{8 \micron_2}$ \\
\hline
linear time$^\dagger$ 	& \ref{eqn:ch2}		& $0.00238 \pm 0.00036$ & $0.00187 \pm 0.00045$ & $0.00307 \pm 0.00055$ & 0   & 0   & 0   \\
quad ln time		& \ref{eqn:quadlnt}	& $0.00271 \pm 0.00042$ & $0.00226 \pm 0.00060$ & $0.00329 \pm 0.00063$ & 14  & 38  & 31  \\
quad time		& \ref{eqn:quadt}	& $0.00291 \pm 0.00059$ & $0.00179 \pm 0.00085$ & $0.00397 \pm 0.00086$ & 16  & 17  & 15  \\
rising exp		& \ref{eqn:risexp}	& $0.00260 \pm 0.00037$ & $0.00188 \pm 0.00047$ & $0.00307 \pm 0.00055$ & 17  & 17  & 17  \\
linear ln time		& \ref{eqn:linlnt}	& $0.00363 \pm 0.00038$ & $0.00318 \pm 0.00046$ & $0.00417 \pm 0.00054$ & 60  & 34  & 25  \\
no time			& \ref{eqn:not}		& $0.00210 \pm 0.00037$ & $0.00103 \pm 0.00046$ & $0.00353 \pm 0.00058$ & 252 & 210 & 18  \\
\hline
\end{tabular}
$^\dagger${For {\sc linear time}: BIC$_{4.5 \micron_{1+2}+8 \micron_{1+2}}=10\,772$, BIC$_{4.5 \micron_{1+2}+8 \micron_1}=8\,298$ and BIC$_{4.5 \micron_{1+2}+8 \micron_2}=8\,072$.}
\end{table*}

\begin{figure}
\includegraphics[width=0.45\textwidth]{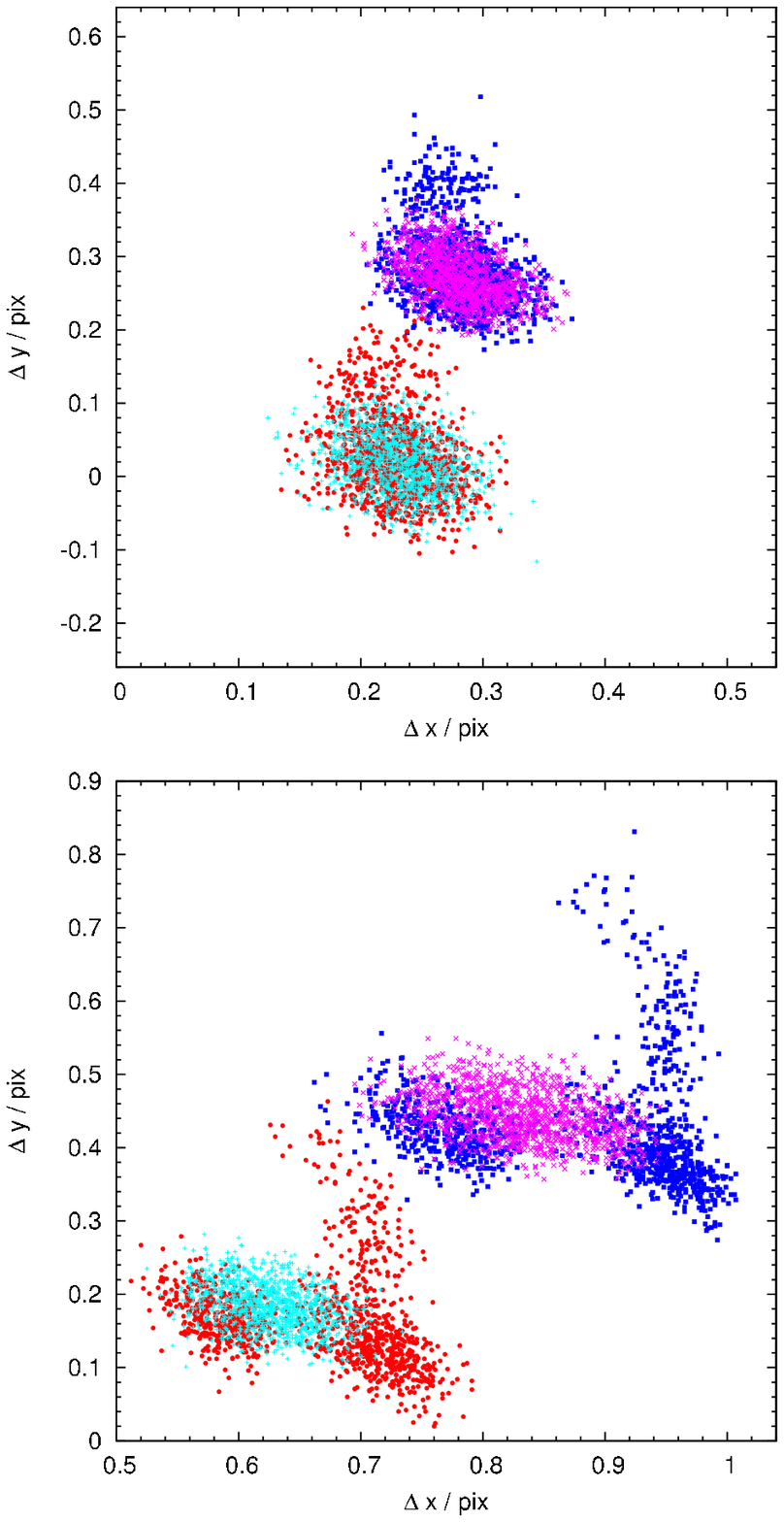}
\caption{The detector positions of WASP-17 both during and outside of 
occultation. 
The top panel shows the occultation of 2009 Apr 24 and the bottom panel 
shows the occultation of 2009 May 1. 
The 4.5 (blue squares) and 8 (red circles) \micron\ data taken outside of 
occultation are overplotted with the 4.5 (magenta saltires) and 8 (cyan crosses) 
\micron\ data taken during occultation.
The 4.5 \micron\ data are shown relative to detector position ($x$,$y$) = 
(24,25), and the 8 \micron\ relative to ($x$,$y$) = (25,23). 
Note that, though the axes' ranges are the same between the two plots, each  
abscissa covers only 60 per cent the range of each ordinate.
\label{fig:detpos}}
\end{figure}

In Figure~\ref{fig:detpos} we present the detector positions of WASP-17 both 
during and outside of occultation. On 2009 Apr 24, the star occupied the same 
region of the detector during the occultation as when outside of occultation. 
However, on 2009 May 1, the star occupied different regions of the detector 
during occultation than when outside of occultation, though there was some 
overlap. 
The reason for this can be seen in the top-right panel of 
Figure~\ref{fig:eclpos}, which shows that the star moved steadily in the 
$x$-direction. This was in addition to the motion due to the nodding of the 
spacecraft, which resulted in some overlap between the in-occultation and 
out-of-occultation detector positions. 
As we decorrelate the light curves with detector position, the data from 2009 
Apr 24, with the greater detector position overlap, are thought to be more 
reliable. However, the data from the two occultations detrend similarly well, 
and we find no reason to disregard the latter dataset.

This indicates that, though we had requested the same detector positions for the 
target for each observation run, small differences in the pointing and motion of 
{\it Spitzer} can result in markedly different systematics. 

We investigated using the fitted PSF positions from the higher signal-to-noise 
4.5 \micron\ data in the aperture photometry and positional decorrelation of the 
8 \micron\ data. 
To account for the offset between the two detectors we fit the differences in 
the $x$ and $y$ directions and translated the coordinates by those amounts. 
The 8 \micron\ occultation depths, both when incorporating one of the two 
datasets or both of them, were very similar to those obtained when fitting the 
stellar PSF position in the 8 \micron\ data, and there was no significant 
reduction in the residual scatter about the best-fitting models. Hence, we 
proceeded as before.

{\it Spitzer}'s pointing oscillates around the nominal position, with an 
amplitude of $\sim$0.1 pixels over a period of $\sim$1 hour. 
We also see higher frequency jitter, with periods of 1--2 minutes (the cadence 
of our data is 12 seconds), in the position of WASP-17. 
Some authors \citep[e.g.][]{2010arXiv1004.0836W} chose to smooth the measured 
target positions prior to light curve detrending. 
However, we found that detrending with the unsmoothed positions resulted in a 
reduced BIC ($\Delta$BIC = $-931$), and in smaller residual RMS values: 5.3 and 
8.8 per cent lower for the two 4.5 \micron\ datasets, and 0.6 and 1.4 per cent 
lower for the two 8 \micron\ datasets. 

To ascertain whether the observed short-period jitter was due to measurement 
error, we measured the position of a second star in the field for the two 4.5 
\micron\ datasets. For both WASP-17 and the second star we subtracted their 
Gaussian-smoothed ($\sigma = 84$ s) positions to remove the longer-period 
oscillations. We then fitted Gaussians to the distributions of the detector $x$ 
and $y$ coordinates of both stars and of their relative separations. 
If the measured positions of WASP-17 and the second star are uncorrelated, then 
the variance of the distribution of relative separations would be the sum of the 
variances of the distributions of each star's positions.
However, we found that the distribution of separation in the $x$-direction had a 
variance smaller than that by a factor nine for the first dataset and by a 
factor two for the second dataset. For the $y$-coordinate, the 
factors were 25 and 6 for the two datasets. Thus, the short-period jitter is 
real and the light curves should be detrended with unsmoothed target positions.

\subsubsection{Aperture radii revisited}
\begin{figure}
\includegraphics{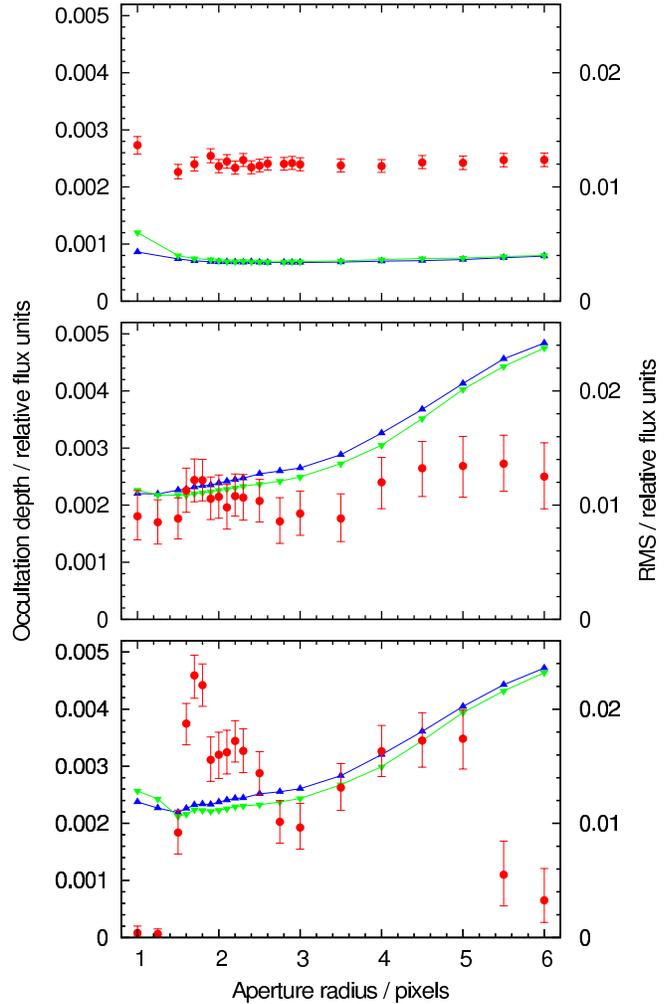}
\caption{
{\bfseries{\em Top panel}}: The dependence on aperture radius of the fitted 
occultation depth (red circles with error bars) and the residuals (blue 
up-triangles = 2009 Apr 24 data, green down-triangles = 2009 May data) for the 
4.5 \micron\ data.  
{\bfseries{\em Middle panel}}: The same as the top panel, but for the 8 \micron\ 
data and when treating the `pixel phase' effect. 
{\bfseries{\em Lower panel}}: The same as the top panel, but for the 8 \micron\ 
data and when neglecting the `pixel phase' effect.
\label{fig:rms-depth}}
\end{figure}

As a check of the choice of aperture radius (2.9 pixels) for the 4.5 \micron\ 
data, we input the 4.5~\micron\ light curves obtained with each aperture radius 
into a simultaneous MCMC analysis that incorporated all but the 8~\micron\ data. 
These analyses produced consistent 4.5~\micron\ occultation depths 
(Figure \ref{fig:rms-depth}, upper panel), indicating that
the 4.5~\micron\ result is relatively insensitive to the choice of aperture 
radius. 

As a check of the choice of aperture radius (1.6 pixels) for the 8 \micron\ 
data, we input each 8~\micron\ light curve into a simultaneous MCMC analysis 
that incorporated all other data. 
When decorrelating with detector position (Figure \ref{fig:rms-depth}, middle 
panel), the fitted 8~\micron\ occultation depth varies weakly with 
aperture radius. 
Beyond an aperture radius of 3.5 pixels (by which point the flux due to the sky 
background is 3 times that of the target within the target aperture), a deeper 
occultation is measured. 
Without decorrelating with detector position (Figure \ref{fig:rms-depth}, lower 
panel), the fitted 8~\micron\ occultation depth is a strong function of aperture 
radius. 

As the conclusions drawn from {\it Spitzer} occultation observations depend on 
accurately measured occultation depths, we advise others to check for a 
correlation between flux and detector position in their 5.8 \micron\ and 
8 \micron\ datasets, and for a dependence of occultation depth on aperture 
radius. 
For example, from Figure~1 of \citet{2010ApJ...711..374F} it appears that 
similar patterns of saw-tooth systematics are present in both the 4.5 and 8 
\micron\ light curve, though they only decorrelate the former light curve with 
detector position. If a dependence of measured flux on detector position was 
introduced during aperture photometry then the measured 8 \micron\ occultation 
depth could be erroneous. 

\subsection{Deconvolution photometry}
\label{sec:decon}
To verify the 8 \micron\ occultation depths and to investigate the source of the 
dependence of the measured 8 \micron\ flux on detector position, we obtained 
8 \micron\ light curves by performing deconvolution photometry with 
{\sc decphot}. 
This method was first described by \citet{decphot1,decphot2} and has been 
optimized for {\it Spitzer} data by Lanotte et al. (in prep). It is based on the  
image-deconvolution method of \citet[][see also \citet{Magain2}]{Magain1}, which 
respects the sampling theorem of \citet{Shannon}, 
in contrast with traditional deconvolution methods. In a first step, 25 random 
BCD images taken on 2009 April 24 were used to determine a 
partial PSF. This was then used to deconvolve the whole set of images and to 
determine optimally the position and flux of WASP-17. 

The {\sc decphot} light curves do not exhibit a position-dependent modulation of the 
flux (Figure~\ref{fig:spitz-ch4}). 
Therefore, the saw-tooth modulation seen in the light curves obtained from 
aperture photometry (Figure~\ref{fig:spitz-ch4}) is likely due to a pixellation 
effect rather than an intra-pixel inhomogeneity in IRAC's 8 \micron\ detector 
response. 
During aperture photometry of the 8 \micron\ data, an aperture radius of only 
1.6 pixels was used. 
The calculation of a circular aperture is non-trivial and the majority of 
photometry routines make a polygonal approximation, which tends to be 
less accurate for smaller radii. 
Aside from that the calculation of how much flux should be attributed to 
partial pixels is another potential source of error. A better result is  
obtained if a PSF is used, rather than if uniform illumination is assumed, but 
even that is not perfect. 
Partial deconvolution is a photometric method that is optimal in a 
least-squares sense, i.e. the background contribution is minimized because each 
pixel is properly weighted. As this is not the case for aperture photometry, and 
as the background at 8 \micron\ is bright relative to the target, we had to use 
a small aperture to optimize the signal-to-noise ratio of our measurements, 
leading to pixelisation effects that translated into a correlation of the 
measured flux with detector position. 

We performed a combined MCMC analysis incorporating the {\sc decphot} 8 \micron\ 
light curves, which were detrended with Equation~\ref{eqn:ch4}. The raw and 
detrended data are shown with the best-fitting trend and occultation models in 
Figure~\ref{fig:spitz-ch4}. 
We found consistent 8 \micron\ occultation depths when incorporating only one 
dataset or both datasets in our analysis (Table~\ref{tab:decphot-depths}). 
The residuals of the {\sc decphot} light curves exhibit a slightly smaller 
scatter than the aperture photometry light curves do 
(Figure~\ref{fig:spitz-ch4}; Table~\ref{tab:decphot-depths}). 
These {\sc decphot} depths and associated uncertainties are in close agreement 
with those derived using the light curves obtained from simple aperture 
photometry (Table~\ref{tab:decphot-depths}). This is also the case for \ecos, 
\esin\ and the time of mid-occultation (Table~\ref{tab:ecc}). 
Thus our method of obtaining 8 \micron\ light curves by simple aperture 
photometry and detrending them with detector position is verified, and it is 
these light curves that we use in the simultaneous analysis from which we 
calculated our system parameter values.

\subsection{Partitioning of data}

In our simultaneous MCMC analysis we partitioned the WASP photometry according 
to observation season and camera into five datasets, so that each dataset could 
thus be normalised independently, as was done in A10. 
As in T10, we partitioned the RV data into four datasets: 
CORALIE data sampling the full orbit ($rv1$); 
HARPS data sampling the full orbit ($rv2$); 
a spectroscopic transit and a week of adjoining data as measured by CORALIE 
($rv3$); 
a spectroscopic transit and two RVs from the following day as measured by 
HARPS ($rv4$). 
Both an instrumental offset and a specific stellar activity level have the 
potential to affect the measured RV of a star. 
The spectroscopic transits comprise a large number of RVs taken in quick 
succession, whereas the data sampling the full orbit were taken over a long 
time span and are thus expected to sample a range of stellar activity level  
that should average to a mean value of zero (T10). 
Thus, by partioning the RV data, we allow each dataset to have its own 
centre-of-mass velocity $\gamma$, thus avoiding the risk of obtaining spurious 
values for the planet's mass and orbital eccentricity. 

\subsection{Photometric and RV noise}

We scaled the photometric error bars so as to obtain a reduced $\chi^{2}$ of 
unity, applying one scale factor per dataset. The aim was to properly 
weight each dataset in the simultaneous MCMC analysis and to obtain realistic 
uncertainties. 
For the five sets of WASP 
photometry the scale factors were in the range 0.87--0.96. The error bars of 
the Euler photometry were multiplied by 1.33. 
The scale factors for the occultation photometry were in the range 1.04--1.09. 
Importantly, the error bars of the occultation photometry were not scaled when 
deciding which trend models or aperture radii to use. 

\begin{figure}
\includegraphics[width=0.45\textwidth]{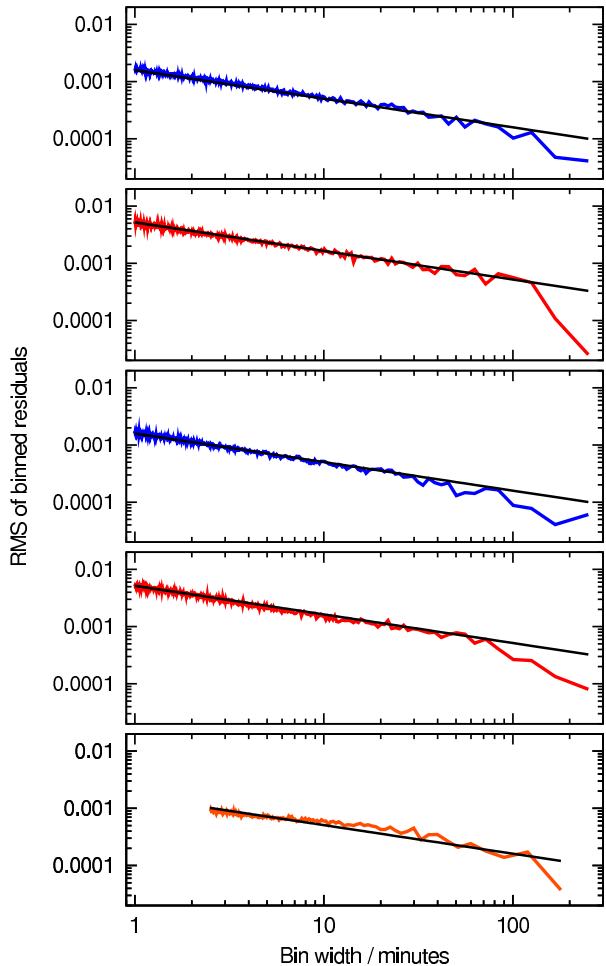}
\caption{RMS of the binned residuals for the new {\it Spitzer} occultation 
photometry (upper four panels, with the datasets presented in the same order as 
in Figure~\ref{fig:spitz}) and the existing Euler photometry (lower panel). 
The solid black lines, which are the RMS of the unbinned 
data scaled by the square root of the number of points in each bin,  show the 
white-noise expectation. 
The ranges of bin widths (1--250 minutes for {\it Spitzer} and 2.5--180 minutes 
for Euler) are appropriate for the datasets' cadences and durations. 
\label{fig:resids}}
\end{figure}

We assessed the presence of correlated noise in the {\it Spitzer} and Euler data 
by plotting the RMS of their binned residuals (Figure~\ref{fig:resids}). 
Though there is no correlated noise evident in the {\it Spitzer} data, it is 
present at a small level in the Euler data over time-scales of 8--80 minutes. 
Due to the similarity with the time-scales of the fitted features in the 
transit (ingress takes 36 minutes, as does egress, and the transit duration is 
264 minutes), the values of some fitted parameters may be affected to a small 
degree. 

For the same reasons as with the photometry, we added a jitter term in 
quadrature to the formal radial velocity errors, as might arise from stellar 
activity. 
We used an initial MCMC run to determine the level of jitter required for each 
dataset to obtain a reduced $\chi^{2}$ of unity. 
We found that the HARPS orbital data ($rv2$) required a jitter of 3 \ms\ and 
the HARPS spectroscopic transit data ($rv4$) required a jitter of 20 \ms. 
It was not necessary to add any jitter to either of the two CORALIE datasets.

\subsection{Time systems and light travel time}
The Euler photometry and the CORALIE RVs are in the BJD (UTC) time system. 
The WASP and {\it Spitzer} photometry are in the HJD (UTC) time system. 
The difference between BJD and HJD is less than 4~s and so is negligible for our 
purposes. 
Although leap second adjustments are made to the UTC system to keep it close to 
mean solar time, meaning one should really use Terrestrial Time, our 
observations span a short baseline (2006--2008), during which there were no leap 
second adjustments. 

The occultation of WASP-17b occurs farther away from us than its transit does, 
so we made a first order correction for the light travel time. We calculated the 
light travel time between the beginning of occultation ingress and the beginning 
of transit ingress to be 50.4 s. We subtracted this from the mid-exposure times 
of the {\it Spitzer} occultation photometry. 
As we measure the time of mid-occultation to a precision of $\pm 150$ s, the 
impact of this correction was small.

\section{Results}
\begin{figure*}
\includegraphics{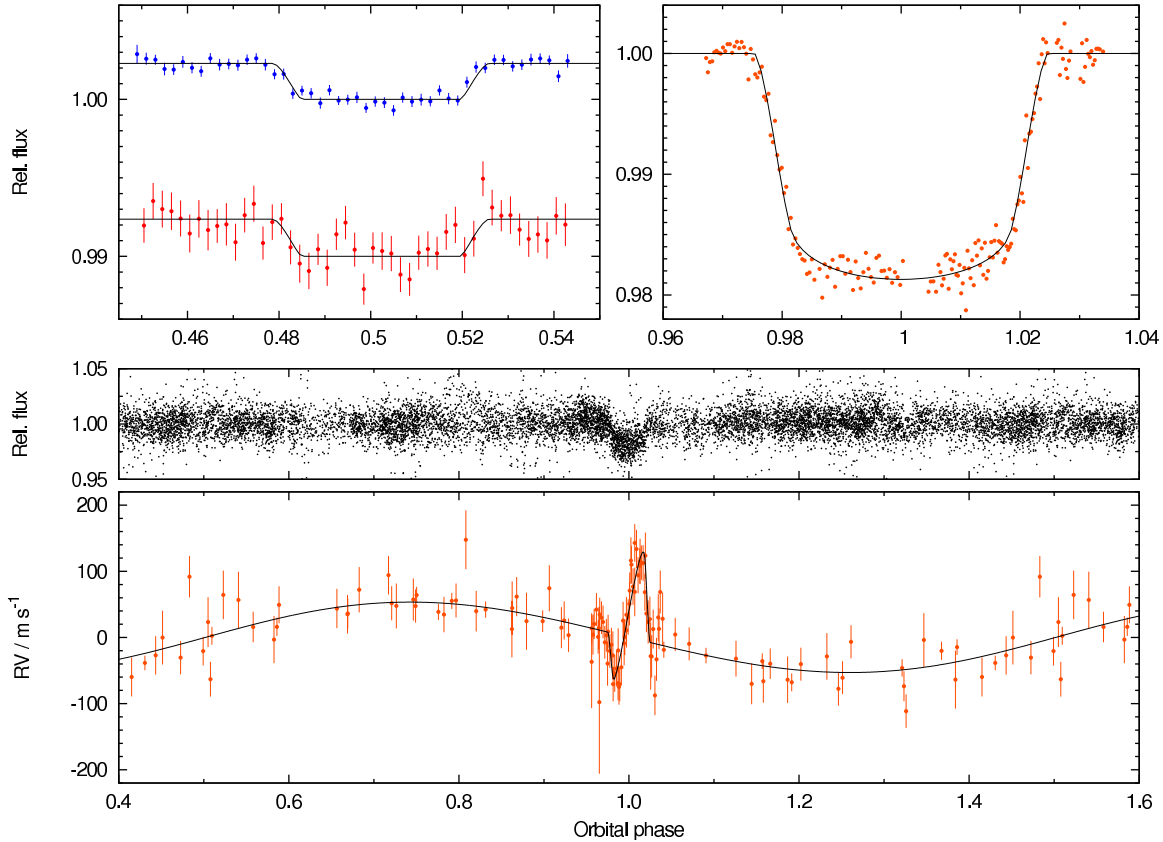}
\caption{
The results of our combined analysis, which combines the new {\it Spitzer} 
occultation photometry with existing photometry and radial velocity measurements. 
The models generated from the best-fitting parameter values of 
Table~\ref{tab:sys-params} are overplotted. 
{\bf\em Top-left}: Occultations at 4.5 \micron\ and, offset in relative flux by 
$-0.01$, 8 \micron. The two occultations per channel from Figure~\ref{fig:spitz} 
were binned ($\Delta \phi=0.002, \sim$11 min) together. 
{\bf\em Top-right}: Transit light curve taken with Euler in the 
{\it I$_{c}$}-band (data from A10). 
{\bf\em Middle}: Photometric orbit and transit illustrated by WASP-South data 
(data from A10). 
{\bf\em Bottom}: Spectroscopic orbit and transit illustrated by CORALIE and 
HARPS data (data from A10 and T10). The measured systemic velocities of each 
dataset (Table~\ref{tab:sys-params}) have been subtracted.
\label{fig:mcmc}}
\end{figure*}

Table~\ref{tab:sys-params} shows the median values and the 1-$\sigma$ 
uncertainties of the fitted proposal parameters and derived parameters from 
our final MCMC analysis. 
Figure~\ref{fig:spitz} shows the best-fitting trend and occultation models 
together with the raw and detrended {\it Spitzer} data. 
Table~\ref{tab:coeffs} gives the best-fitting values for the parameters of the 
trend models (Equation~\ref{eqn:ch2}), together with their 1 $\sigma$ 
uncertainties.  
Figure~\ref{fig:mcmc} displays all the photometry and RVs used in the MCMC 
analysis, with the best-fitting eclipse and radial velocity models superimposed.

\begin{table*} 
\caption{System parameters of WASP-17} 
\label{tab:sys-params} 
\begin{tabular}{llcl}
\hline 
Parameter & Symbol & Value & Unit\\ 
\hline 
\\
Orbital period & $P$ & $3.7354380 \pm 0.0000068$ & d\\
Epoch of mid-transit (HJD, UTC) & $T_{\rm c}$ & $2454577.85806 \pm 0.00027$ & d\\
Transit duration & $T_{\rm 14}$ & $0.1830 \pm 0.0017$ & d\\
Duration of transit ingress $\approx$ duration of transit egress & $T_{\rm 12} \approx T_{\rm 34}$ & $0.0247 \pm 0.0017$ & d\\
Planet/star area ratio & (\rplanet/\rstar)$^2$ & $0.01696 \pm 0.00026$ & \\
Impact parameter & $b$ & 0.401$^{+ 0.059}_{- 0.077}$ & \\
Orbital inclination & $i$ & 86.83$^{+ 0.68}_{- 0.56}$ & $^\circ$\\
\noalign{\medskip}
Stellar radial reflex velocity & $K_{\rm 1}$ & $53.2 \pm 3.4$ & \ms\\
Semi-major axis & $a$ & $0.05150 \pm 0.00034$ & AU\\
Centre-of-mass velocity & $\gamma_{\rm rv1}$ & $-49\,513.67 \pm 0.56$ & \ms\\
Offset between RV dataset rv2 and rv1 & $\gamma_{\rm rv2-rv1}$ & $22.07 \pm 0.68$ & \ms\\
Offset between RV dataset rv3 and rv1 & $\gamma_{\rm rv3-rv1}$ & $13.5 \pm 2.2$ & \ms\\
Offset between RV dataset rv4 and rv1 & $\gamma_{\rm rv4-rv1}$ & $25.6 \pm 2.8$ & \ms\\
\noalign{\medskip}
Orbital eccentricity & $e$ & 0.028$^{+ 0.015}_{- 0.018}$ & \\
& & $0.0019 < e < 0.0701$ (3 $\sigma$) & \\
Expectation value of orbital eccentricity & $<$$e$$>$ & 0.0055 & \\
Argument of periastron & $\omega$ & $-82.6^{+ 14.6}_{- 2.6}$ & $^\circ$\\
& $e\cos\omega$ & 0.00352$^{+ 0.00076}_{- 0.00073}$ & \\
& $e\sin\omega$ & $-0.027^{+ 0.019}_{- 0.015}$ & \\
Phase of mid-occultaion, having accounted for light travel time & $\phi_{\rm mid-occ.}$ & $0.50224 \pm 0.00050$ & \\
Occultation duration & $T_{\rm 58}$ & 0.1746$^{+ 0.0056}_{- 0.0042}$ & d\\
Duration of occultation ingress $\approx$ duration of occultation egress & $T_{\rm 56} \approx T_{\rm 78}$ \medskip & $0.0232 \pm 0.0016$ & d\\
\noalign{\medskip}
Relative planet-star flux at 4.5 \micron & \fctwo & $0.00229 \pm 0.00013$ & \\
Relative planet-star flux at 8 \micron & \fceight & $0.00237 \pm 0.00039$ & \\
Planet brightness temperature$^{\dagger}$ at 4.5 \micron & $T_{\rm b,4.5\micron}$ & $1881 \pm 50$ & K\\
Planet brightness temperature$^{\dagger}$ at 8 \micron & $T_{\rm b,8\micron}$ \medskip & $1580 \pm 150$ & K\\
\noalign{\medskip}
Sky-projected stellar rotation velocity & \vsini & 10.05$^{+ 0.88}_{- 0.79}$ & \kms\\
Sky-projected angle between stellar spin and planetary orbit axes & $\lambda$ & $-148.7^{+ 7.7}_{- 6.7}$ & $^\circ$\\
\noalign{\medskip}
Star mass & \mstar & $1.306 \pm 0.026$ & \msol\\
Star radius & \rstar & $1.572 \pm 0.056$ & \rsol\\
Star density & \densstar & $0.336 \pm 0.030$ & \denssol\\
Star surface gravity & $\log g_{*}$ & $4.161 \pm 0.026$ & (cgs)\\
Star effective temperature & \teff\ & $6650 \pm 80$ & K\\
Star metallicity & \feh & $-0.19 \pm 0.09$ & \\
\noalign{\medskip}
Planet mass & \mplanet & $0.486 \pm 0.032$ & \mjup\\
Planet radius & \rplanet & $1.991 \pm 0.081$ & \rjup\\
Planet density & \densplanet & $0.0616 \pm 0.0080$ & \densjup\\
Planet surface gravity & $\log g_{\rm P}$ & $2.448 \pm 0.042$ & (cgs)\\
Planet equilibrium temperature$^{\ddagger}$ (full redistribution) & \teqlfull & $1771 \pm 35$ & K\\
Planet equilibrium temperature$^{\ddagger}$ (day side redistribution) & \teqlday & $2106 \pm 41$ & K\\
\\ 
\hline 
\multicolumn{4}{l}{$^{\dagger}$ We modelled both star and planet as black bodies 
and took account of only the occultation depth uncertainty, which dominates.}\\
\multicolumn{4}{l}{$^{\ddagger}$ 
\teqlf\ $= f^{\frac{1}{4}}T_{\rm eff}\sqrt{\frac{R_*}{2a}}$ where $f$ is the 
redistribution factor, with $f=1$ for full redistribution and $f=2$ for day 
side redistribution.}\\
\multicolumn{4}{l}{We assumed the planet albedo to be zero, $A=0$.}
\end{tabular} 
\end{table*}

From this we see that 
WASP-17b is a very bloated planet (\rplanet\ = 2.0 \rjup) in a slightly 
eccentric, 3.7 day, retrograde orbit around an F6V star. 
By constraining the eccentricity of WASP-17b's orbit to low values we have 
shown that the circular solution presented in A10 (in which a total of three 
solutions were presented) is closest to reality. 

\subsection{Orbital eccentricity}

\begin{figure}
\includegraphics[width=8.4cm]{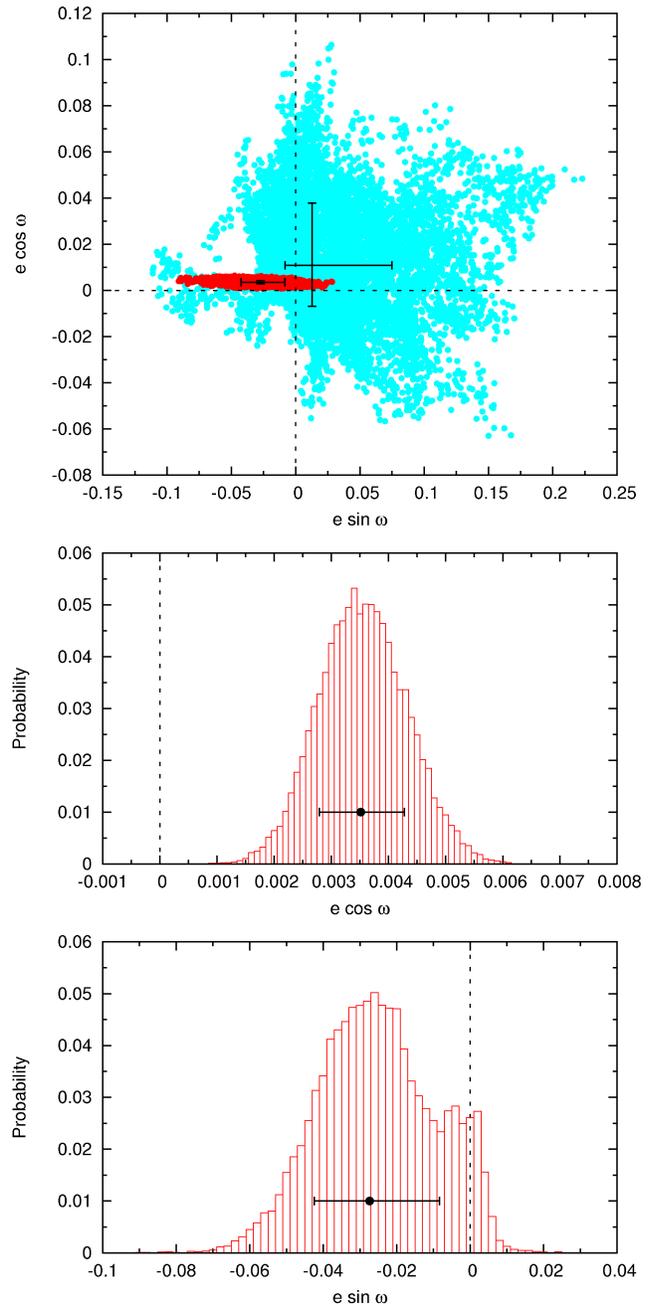}
\caption{{\bfseries{\em Top panel}}: A comparison of the posterior probability 
distributions of \ecos\ and \esin\ from our combined MCMC analysis when 
including (red dots) and excluding (cyan dots) the occultation photometry. 
The extent of the error bars show the 1-$\sigma$ confidence limits and their 
intersections show the median values. 
{\bfseries{\em Middle panel}}: Normalised histogram of the \ecos\ posterior 
probability distribution from our combined MCMC analysis incoroporating the 
{\it Spitzer} photometry. The point with error bars, arbitrarily placed at 
probability = 0.01, 
depicts the best-fitting value and its 1-$\sigma$ error bars.
{\bfseries{\em Bottom panel}}: The same plot as the middle panel, but for \esin.
\label{fig:ecos-esin}}
\end{figure}

\begin{figure}
\includegraphics[width=8.4cm]{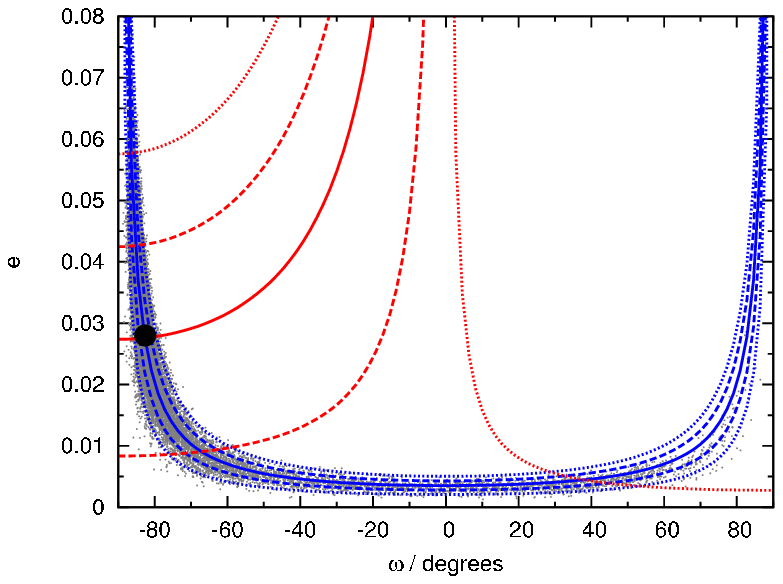}
\caption{The range of $e$ and $\omega$ permitted by the available data. 
The black dot with error bars shows the best-fitting values from our combined 
MCMC analysis. 
The grey dots are the values in accepted MCMC steps. 
The solid blue and red lines show the values of $e$ and $\omega$ that would be 
indicated by, respectively, the best-fitting values of \ecos\ and \esin\ on 
their own. 
The dashed and the dotted lines bound the parameter space permitted by the 
1-$\sigma$ and the 2-$\sigma$ limits, respectively, on those parameters, with 
the same colour scheme applying. 
Note that, as the 2-$\sigma$ upper limit on \esin\ is positive, almost all 
values of $\omega$ are consistent with the data at the 2-$\sigma$ level 
(providing $e < 0.01$).
\label{fig:e_omega}}
\end{figure}

We have shown the orbit of WASP-17b to be non-circular: 
\ecos\ is non-zero at the 4.8-$\sigma$ level 
($\ecos = 0.00352^{+0.00076}_{-0.00073}$; Figure~\ref{fig:ecos-esin}), and the 
best-fitting solution suggests that WASP-17b is occulted by its host star 
$12.0 \pm 2.5$~min later than if it were in a circular orbit. 
Our measurement of \ecos\ rules out large values of $e$ for all orbital 
orientations other than those with $|\omega| \approx 90$, and 
the limits we place on \esin\ prohibits 
large values of $e$ for those orientations with $|\omega| \approx 90$ 
(Figure~\ref{fig:e_omega}).
From the MCMC analysis, the 1-$\sigma$ (68.3 per cent) lower and upper limits on 
$e$ are, respectively, 0.010 and 0.043 and the 3-$\sigma$ (99.7 per cent) lower 
and upper limits on $e$ are, respectively, 0.0019 and 0.0701. 
We can set a more stringent 3 $\sigma$ lower limit on $e$ 
by assuming $\esin \approx 0$ (and so $|\omega| \approx 90$), in which case it 
would be equal to that of the 3-$\sigma$ lower limit on \ecos: 0.0012. 

Almost all values of $\omega$ are permitted by the current data, with only 
$|\omega| \approx 90$ being ruled out by the limits placed on \esin\ 
(Figure \ref{fig:e_omega}). 
Large values of $e$ are consistent with the data only if $|\omega| \approx 90$,  
otherwise any orientation of the orbital major axis is permitted providing that 
$e$ is small. 
We can thus use our measurement of \ecos\ to infer a {\it probable} value of 
$e$. 
For random orientations of the major axis, the {\it expected value} of 
$\cos \omega$ is $\E(\cos \omega) = 2/\pi$. Thus, the {\it expected value} of 
$e$ is $\E(e) = \ecos / \E(\cos \omega) = 0.0055$. 

We explored the effect of each occultation photometry dataset in turn on the 
orbital eccentricity, and of all four datasets combined. 
We did so by performing MCMC runs that incoporated either all, none or just one 
of the {\it Spitzer} datasets (Table~\ref{tab:ecc}; Figure~\ref{fig:ecos-esin}). 
This demonstrates how valuable the {\it Spitzer} occultation photometry is in 
determining orbital eccentricity, as its inclusion in our combined analysis 
caused the size of the 68.3 per cent confidence interval for \ecos\ to 
decrease by a factor 30.2, and the interval for \esin\ to decrease in size by a 
factor 2.4.
In addition to the RV data, it is the orbital phase of the occultation that 
constrains \ecos\ and it is the occultation duration, relative to the transit 
duration, that constrains \esin\ \citep{2005ApJ...626..523C}. 
When including any one of the four occultation datasets, the best-fitting values 
of \ecos\ and \esin\ obtained are consistent with the values obtained 
when including all four datasets. Thus no individual dataset is biasing our 
best-fitting solution. 

\begin{table*}
\caption{Effect of occultation light curves on best-fitting orbital eccentricity} 
\label{tab:ecc} 
\begin{tabular}{llllll}
\hline
Included occultation photometry	& $e$				 & $\omega$ $(^{\circ})$	& \ecos					& \esin				  & $T_{\rm occ}-T_{\rm occ,circular}$ (min)$^{\dagger}$ \\
\hline
\smallskip
4.5 \micron, 2009 Apr 24	& $0.052^{+ 0.017}_{- 0.020}$	 & $-85.9^{+ 2.7}_{- 1.2}$	& $0.00371^{+ 0.00085}_{- 0.00086}$	& $-0.051^{+ 0.020}_{- 0.017}$	  & ~$12.7 \pm 2.9$		\\
\smallskip
4.5 \micron, 2009 May  1	& $0.0055^{+ 0.0075}_{- 0.0024}$ & $-13^{+ 82}_{- 56}$		& $0.00302^{+ 0.00103}_{- 0.00098}$	& $-0.001^{+ 0.008}_{- 0.007}$	  & ~$10.3^{+3.5}_{-3.4}$	\\
\smallskip
8 \micron,   2009 Apr 24	& $0.015^{+ 0.059}_{- 0.012}$	 & $-92^{+ 184}_{- 21}$		& $0.0021^{+ 0.0039}_{- 0.0067}$	& $-0.005^{+ 0.011}_{- 0.062}$	  & $-7.3^{+13.2}_{-23.2}$	\\
\smallskip
8 \micron,   2009 May  1	& $0.049^{+ 0.020}_{- 0.024}$	 & $-82.2^{+ 7.1}_{- 2.4}$	& $0.00662^{+ 0.00099}_{- 0.00111}$	& $-0.049^{+ 0.024}_{- 0.020}$	  & ~$22.7^{+3.4}_{-3.8}$	\\
\smallskip
None				& $0.038^{+ 0.045}_{- 0.026}$	 & ~\,~$52.6^{+ 14.6}_{- 2.6}$	& $0.011^{+ 0.027}_{- 0.018}$		& ~\,~$0.013^{+ 0.062}_{- 0.021}$ & ~$37^{+92}_{-61}$		\\
\smallskip
All				& $0.028^{+ 0.015}_{- 0.018}$	 & $-82.6^{+ 14.6}_{- 2.6}$	& $0.00352^{+ 0.00076}_{- 0.00073}$	& $-0.027^{+ 0.019}_{- 0.015}$	  & ~$12.0 \pm 2.5$		\\
All (decon. phot.)		& $0.022^{+ 0.016}_{- 0.016}$	 & $-81.2^{+ 27.4}_{- 3.7}$	& $0.00335^{+ 0.00073}_{- 0.00075}$	& $-0.022^{+ 0.017}_{- 0.016}$	  & ~$11.5 \pm 2.6$		\\
\hline
\multicolumn{6}{l}{$^{\dagger}$ $T_{\rm occ}$ is the time of mid-occultation 
derived from a simultaneous MCMC analysis.}\\
\end{tabular}
\end{table*}

\section{Discussion}

\subsection{Planet radius}
With a radius of 2.0 \rjup, WASP-17b is the largest known planet by a margin of 
0.2 \rjup, and is over 0.7 \rjup\ larger than predicted by standard 
cooling theory of irradiated gas giant planets \citep{2007ApJ...659.1661F}.

\citet{2009ApJ...700.1921I} and \citet{2011ApJ...727...75I} used a coupled 
radius-orbit evolutionary model to show that planet radii can be inflated to 
2 \rjup\ and beyond during a transient phase of heating caused by tidal 
circularisation of a short ($a \approx 0.1$), highly eccentric ($e \approx 0.8$) 
orbit. 
Though, as was noted in both studies, planets can persist in an inflated state 
for Gyr beyond the circularisation of their orbit and the cessation of tidal 
heating, they do cool and contract significantly prior to full circularisation. 
In each study the orbits are still significantly non-circular ($e \gtrsim 0.1$) 
when the planets are largest. 
Thus, under the transient heating scenario, the very largest planets are 
expected to have a non-zero eccentricity. 
Though we do measure a non-zero eccentricity for WASP-17b, it is small, and the 
stringent upper limit that we place on $e$ is inconsistent with current models 
of one transient phase of tidal heating. 

Other than transient heating, ongoing tidal heating may occur if 
the orbit of a planet were kept non-circular by the continuing 
interaction with a third body \citep{2010ApJ...713..751I}. 
However, the stringent upper limit we place on $e$ makes this unlikely as the 
sole cause of the inflation of WASP-17b, as it would necessitate a lower 
planetary tidal dissipation factor than theoretical models or empirical 
determinations generally suggest \citep{2010ApJ...713..751I}. 

If the atmospheric opacity of WASP-17b were enhanced then its internal 
heat would be lost at a lower rate and contraction would be slowed 
\citep{2007ApJ...661..502B}. 
The atmospheric opacities of WASP-17b may be enchaned if, for example, the 
strong optical and UV irradiation of the planet by its host star produces 
thick hazes, absorbing clouds and non-equilibrium chemical species 
(e.g. tholins or polyacetylenes).

The bloated planets are all very strongly irradiated by their host stars, and a 
small fraction of stellar insolation energy would be sufficient to account for 
the observed degrees of bloating.
\citet{2002A&A...385..156G} suggested that the kinetic energy of strong winds, 
induced in the atmosphere of a short-period planet by the large day-night 
temperature contrasts that result from tidal locking, may be transported 
downward and deposited as thermal energy in the deep interior. 
However, a mechanism to convert the kinetic energy into thermal energy would 
still be required. 
\citet{2010ApJ...725.1146L} and \citet{2010ApJ...721.1113Y} found that 
turbulence is efficient at dissipating kinetic energy. 
Magnetic drag on weakly ionized winds \citep{2010ApJ...719.1421P} and Ohmic 
heating \citep{2010ApJ...714L.238B} are alternative mechanisms.

\subsection{Planetary atmosphere}

\citet{2008ApJ...678.1419F} hypothesise that the presence of high opacity TiO 
and VO gases in the atmospheres of highly irradiated planets (those experiencing 
an incident flux of $>10^9$ erg s$^{-1}$ cm$^{-2}$) cause them to have 
temperature inversions. Thus, with an incident flux of $2.2 \pm 0.2 \times 
10^{9}$ erg s$^{-1}$ cm$^{-2}$, WASP-17b is expected to have an atmospheric 
temperature inversion under this hypothesis. 

However, \citet{2009ApJ...699.1487S} suggest that, for a planet with the 
insolation level of WASP-17b, it is unlikely that a temperature inversion could 
be caused by the presence of TiO and VO in the upper atmosphere. 
They find that a cold trap exists between the hot convection zone 
and the hot upper atmosphere on the irradiated day side, in which titanium is 
likely to form condensates that settle more strongly 
than does gasesous TiO. Therefore, unless there is extremely vigorous 
macroscopic mixing and the condensed Ti is lofted back in to the upper 
atmosphere then it is unlikely that TiO can explain the observed temperature 
inversions. 
Not only does VO have the same `cold trap' issue, but it also has a lower 
opacity than TiO and is an order of magnitude less abundant.

\citet{2010ApJ...720.1569K} suggest that planets orbiting chromospherically 
active stars do not have temperature inversions, and planets orbiting quieter 
stars do have inversions. 
They suggest that the high UV flux that planets orbiting active stars 
are likely to experience destroys the compounds responsible for the observed 
temperature inversions. 
\citet{2010ApJ...720.1569K} find the two classes to be delineated by a host-star 
activity level of $\log(R^{'}_{\rm HK}) \approx -4.9$. Though they caution that 
the calibration for $\log(R^{'}_{\rm HK})$ is uncertain for stars as hot as 
WASP-17, they measure the star to be quiet: $\log(R^{'}_{\rm HK}) = -5.3$. 
WASP-17b is therefore expected to have an atmospheric temperature inversion 
under this hypothesis as well. 

\begin{figure}
\includegraphics[width=9cm]{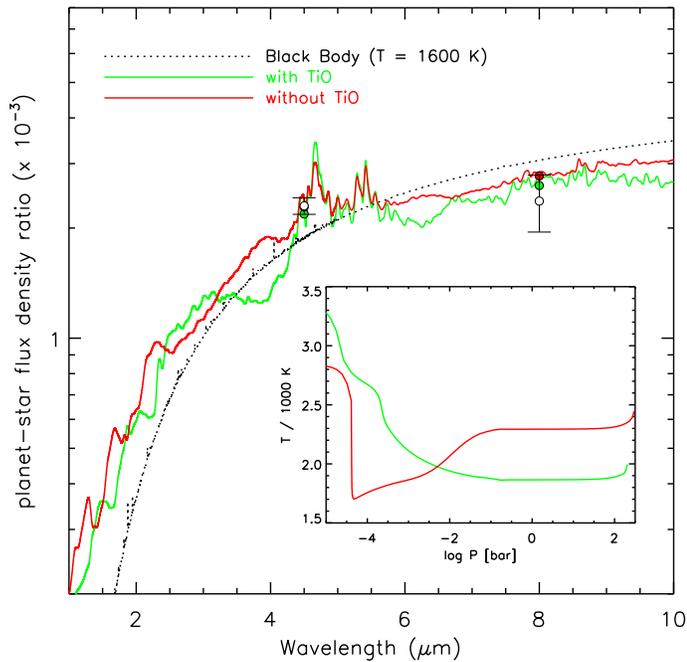}
\caption{Comparison of planet-star flux density measurements with two model 
planet atmospheres and with a black-body. 
The model atmosphere with TiO exhibits a temperature inversion that extends 
down to photospheric depths, whilst the model without TiO does not.
{\bf {\em Inset}}: Temperature-pressure profiles for the two model atmospheres. 
\label{fig:atmos}}
\end{figure}

In Figure~\ref{fig:atmos}, the measured 4.5 and 8 \micron\ planet-star 
flux-density ratios are compared to two model atmosphere spectra of the planet 
\citep{2005ApJ...632.1132B}, with parameters taken from 
Table~\ref{tab:sys-params}. 
A black-body (\teql = 1600 K) is a poor fit to the data and is thus ruled out. 
In one model atmosphere TiO produces a temperature inversion across the 
photospheric depths. In the other model, there is no atmospheric TiO.
The two models have near-identical 4.5 and 8 \micron\ absolute fluxes, and 
so we can not currently discriminate between the two. 
A precise measurement at 3.6 \micron\ may distinguish between the two cases and 
thus reveal whether WASP-17b has an atmospheric temperature inversion. 

By modelling the planet and star as black bodies, we used the measured 
planet-star flux-density ratios to calculate 4.5 and 8 \micron\ brightness 
temperatures of $1881 \pm 50$ K and $1580 \pm 150$ K, respectively. 
We calculate an equilibrium temperature \teqlfull\ = $1771 \pm 35$ K by 
modelling the planet as a black body with efficient redistribution of energy 
from its day side to its night side. The closeness of the brightness 
temperatures to this equilibrium temperature is consistent with the planet 
having a low albedo and efficient heat redistribution. 

\subsection{Misaligned orbit}
WASP-17b is in a retrograde orbit. For planet-planet or star-planet scattering 
to have caused the misalignment between the orbit of WASP-17b and the spin axis 
of its host star, an additional body must have been present. 
We looked for evidence of a long-term drift $\dot\gamma$ in the radial velocity 
measurements, which span 716 days, as may be caused by the presence of a 
long-period companion. 
From a straight-line fit to the residuals of the radial velocities about 
the best-fitting model, we get $\dot\gamma = -6 \pm 5$ \ms\ yr$^{-1}$. Hence, 
there is currently no evidence for a third body in the system, but this does not 
preclude planet-planet scattering as the cause of the misalignment.
\citet{2008ApJ...678..498N} found, whilst showing that a combination of 
planet-planet scattering and the Kozai mechanism can put planets into short, 
retrograde orbits, that the outer planets can end up at large orbital distances, 
making them difficult to detect, or they can be ejected from the system. 

\subsection{System age}
\begin{figure}
\centering                     
\includegraphics[width=84mm]{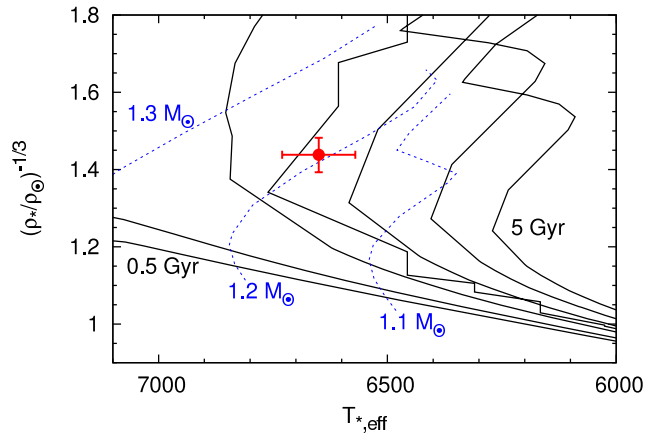}
\caption{Modified H-R diagram. 
The isochrones ($Z = 0.012\approx$ \feh\ = $-$0.19) for the ages 0.5, 1, 2, 2.5, 
3, 4, and 5 Gyr are from \citet{2008A&A...482..883M} and the evolutionary mass 
tracks ($Z = 0.012 \approx$ \feh\ = $-$0.019; $Y = 0.30$) are from 
\citet{2008A&A...484..815B}.
To obtain the mass tracks, we performed a simple linear interpolation of their  
$Z = 0.0008$ and $Z = 0.017$ tracks.
\label{fig:evol}}
\end{figure}

We interpolated the stellar evolution tracks of \citet{2008A&A...482..883M} 
using \densstar\ from Table~\ref{tab:sys-params} and the values of \teff\ 
and \feh\ from T10 (Figure~\ref{fig:evol}). This suggests an 
age of $2.65 \pm 0.25$ Gyr and a mass of $1.20 \pm 0.05$ \msol\ for WASP-17.

Assuming the stellar-spin axis to be in the sky plane, the measured \vsini\ of 
WASP-17 and its derived stellar radius (Table~\ref{tab:sys-params}) indicate an 
upper limit to the rotational period of $P_{\rm rot} = 7.91 \pm 0.75$ d. 
Combining this with the $B - V$ colour of an F6V star from 
\citet{2008oasp.book.....G}, and the relationship of 
\citet{2007ApJ...669.1167B}, we estimate an upper limit on the gyrochronological 
age of $1.9 \pm 0.5$ Gyr. We found no evidence for rotational modulation in the 
WASP light curves. 

We calculated a tidal circularisation time-scale of $\tau_{\rm circ} = 5$ Myr 
for WASP-17b by using the best-fitting values of the planetary 
($Q_{\rm P} = 10^{5.5}$) and stellar ($Q_{*} = 10^{6.5}$) tidal dissipation 
factors of \citet{2008ApJ...678.1396J} in their Equation 1. As the values of the 
tidal dissipation factor are highly uncertain 
\citep[$Q_{\rm P} = 10^5$--$10^8$, $Q_{*} = 10^5$--$10^8$, e.g.][]
{2011ApJ...727...75I}, a range of $\tau_{\rm circ} =$ 2--1700 Myr is possible.

With \teff\ = $6650 \pm 80$ K (T10), WASP-17 is in the `Lithium gap' (or `dip'), 
which is the range of \teff\ = $6600 \pm 150$ K in which stars are depleted in 
lithium by a factor of 30 or more than in hotter and cooler stars 
\citep[see][and references therein]{1995ApJ...446..203B}. 
The upper limit placed on the lithium abundance (\ali\ $< 1.3$) in A10 is 
consistent with this. Thus, lithium is not an effective indicator of age for 
WASP-17.

\section{Conclusions}

\subsection{Science}

With a radius of 2.0 \rjup, WASP-17b is larger than any other known planet by 
0.2 \rjup\, and it is 0.7 \rjup\ larger than predicted by standard cooling 
theory of irradiated gas giant planets. 
The extent of the planet's inflation is difficult to explain with current 
models. 

Our {\it Spitzer} occultation photometry gives much 
tighter constraints on orbital eccentricity than existing radial 
velocity data alone, thus permitting an accurate determination of the stellar 
and planetary radii. 
We have shown that WASP-17b is in a slightly eccentric orbit, with 
$0.0017 < e < 0.0701$. 
The stringent upper limit we have placed on eccentricity suggests that a 
transient phase of tidal heating alone could not have inflated the planet to 
its measured radius.
Nor could ongoing tidal heating involving a third body, unless the planetary 
tidal quality factor is smaller than the best theoretical and empirical 
determinations. 

We find no evidence in the radial velocity measurements for a third body in the 
system, the presence of which would be necessary to excite the eccentricity of 
WASP-17b for tidal heating to be ongoing, and may have been necessary to 
misalign the planet's orbital axis with the spin axis of the star. 

Our 4.5 and 8 \micron\ planet-star flux-density ratios do not probe the 
existence of the expected atmospheric temperature inversion, but a measurement 
at 3.6 \micron\ may do so. 
Though the ratios are inconsistent with a black-body atmosphere, they are 
consistent with a low-albedo planet that efficiently redistributes heat from 
its day side to its night side. 

\subsection{{\it Spitzer} data}

To determine correctly the photometric uncertainties and the optimal 
aperture radii to use for {\it Spitzer} data, account must be taken of the 
counts removed during sky-dark subtraction. 

When the background is bright relative to the target at 8 \micron, the measured 
occultation depth can depend sensitively on the choice of aperture radius. In 
these circumstances detrending with detector position vastly reduces the 
dependency. An alternative is to perform deconvolution photometry. 

In addition to the known hour-long oscillations of {\it Spitzer}'s pointing 
about the nominal position, there is also a high-frequency jitter, with periods 
of 1--2 minutes. 
So, when accounting for the inhomogeneous detector response (or `pixel phase' 
effect), one should detrend target flux with the unsmoothed target detector 
positions.

\section*{Acknowledgments}
This work is based in part on observations made with the Spitzer 
Space Telescope, which is operated by the Jet Propulsion Laboratory, 
California Institute of Technology under a contract with NASA. 
Support for this work was provided by NASA through an award issued by 
JPL/Caltech. 
M. Gillon acknowledges support from the Belgian Science Policy Office in the 
form of a Return Grant. 


\label{lastpage}

\end{document}